\documentclass[a4paper]{ar-1col}
\usepackage{graphicx,amsmath,amssymb,bm}
\usepackage{color}
\usepackage{array}
\jyear{2015}

\newcommand{\be}{\begin{equation}}	
\newcommand{\ee}{\end{equation}}
\newcommand{\vlowk}{V_{{\rm low}\,k}}
\newcommand{\tneff}{{\rm 3N_{eff}}}

\newcommand{\mev}{\, \text{MeV}}
\newcommand{\kev}{\, \text{keV}}

\newcommand{\hw}{\hbar\omega}
\newcommand{\pfg}{pfg_{9/2}}
\newcommand{\sdfp}{sdf_{7/2}p_{3/2}}

\newcommand{\dt}{d_{3/2}}
\newcommand{\so}{s_{1/2}}

\newcommand{\gn}{g_{9/2}}

\begin{document}

\markboth{K.\ Hebeler, J.\ D.\ Holt, J.\ Men\'endez and A.\ Schwenk}%
{Neutron-rich nuclei and neutron-rich matter}

\title{Nuclear forces and \\ their impact on \\
neutron-rich nuclei and neutron-rich matter}

\author{K.\ Hebeler,$^{1,2}$ J.\ D.\ Holt,$^{1,2,3}$ J.\ Men\'endez$^{1,2,4}$
and A.\ Schwenk$^{1,2}$
\affil{$^1$Institut f\"ur Kernphysik, Technische Universit\"at
Darmstadt, 64289 Darmstadt, Germany}
\affil{$^2$ExtreMe Matter Institute EMMI, 
GSI Helmholtzzentrum f\"ur Schwerionenforschung GmbH, 
64291 Darmstadt, Germany}
\affil{$^3$TRIUMF, 4004 Wesbrook Mall, Vancouver, British Columbia, 
V6T 2A3 Canada}
\affil{$^4$Department of Physics, University of Tokyo, Hongo, Tokyo 
113-0033, Japan}
\affil{emails: kai.hebeler@physik.tu-darmstadt.de, jholt@triumf.ca,
menendez@nt.phys.s.u-tokyo.ac.jp, schwenk@physik.tu-darmstadt.de}}

\begin{abstract}
We review the impact of nuclear forces on matter at neutron-rich
extremes. Recent results have shown that neutron-rich nuclei become
increasingly sensitive to three-nucleon forces, which are at the
forefront of theoretical developments based on effective field
theories of quantum chromodynamics. This includes the formation of
shell structure, the spectroscopy of exotic nuclei, and the location
of the neutron dripline. Nuclear forces also constrain the properties
of neutron-rich matter, including the neutron skin, the symmetry
energy, and the structure of neutron stars. We first review our
understanding of three-nucleon forces and show how chiral effective
field theory makes unique predictions for many-body forces. Then, we
survey results with three-nucleon forces in neutron-rich oxygen and
calcium isotopes and neutron-rich matter, which have been explored
with a range of many-body methods. Three-nucleon forces therefore
provide an exciting link between theoretical, experimental and
observational nuclear physics frontiers.
\end{abstract}

\begin{keywords}
three-nucleon forces, exotic nuclei, neutron-rich matter, neutron stars
\end{keywords}
\maketitle

\tableofcontents

\section{INTRODUCTION}
\label{intro}

Recent studies have shown that three-nucleon (3N) forces play a key
role for understanding and predicting neutron-rich nuclei and for the
formation and evolution of shell structure. In addition, the location
of the neutron dripline, where nuclei cease to be bound, is sensitive
to small interaction contributions and therefore to 3N forces. At the
same time, 3N forces are the dominant uncertainty in constraining the
properties of neutron-rich matter at nuclear densities, which is important
for the structure of neutron stars. This leads to an exciting connection
of 3N forces with the exploration of extreme neutron-rich nuclei at
rare isotope beam facilities and with forefront observations in
astrophysics.

In the following, we discuss our understanding of nuclear forces based on
chiral effective field theory (EFT) and show how this framework makes unique
predictions for many-body forces. In particular, the properties of all nuclei
can be predicted up to high order (next-to-next-to-next-to-leading order,
N$^3$LO), with just two new low-energy couplings in many-body forces. To this
order all other three- and four-nucleon (4N) interactions depend on couplings
that are determined from interactions with pions or enter in two-nucleon (NN)
interactions. We discuss theoretical developments of the subleading 3N forces
and advances towards including them in few- and many-body calculations. In
Sections~\ref{oxygen} and~\ref{calcium}, we survey results with 3N forces in
neutron-rich oxygen and calcium isotopes, which have been explored with a
range of many-body methods. These present key regions in the study of the
neutron dripline, of shell structure, and in the spectroscopy of exotic
nuclei. In Section~\ref{matter}, we discuss the impact of nuclear forces on
the properties of neutron-rich matter, including the neutron skin, the
symmetry energy, and the properties of neutron stars. Finally, we conclude and
give an outlook with open problems and opportunities in Section~\ref{summary}.

\subsection{Chiral EFT for nuclear forces}
\label{forces}

\begin{figure}[t]
\begin{center}
\includegraphics[trim=5mm 11mm 127mm 21mm,width=0.55\textwidth,clip=]%
{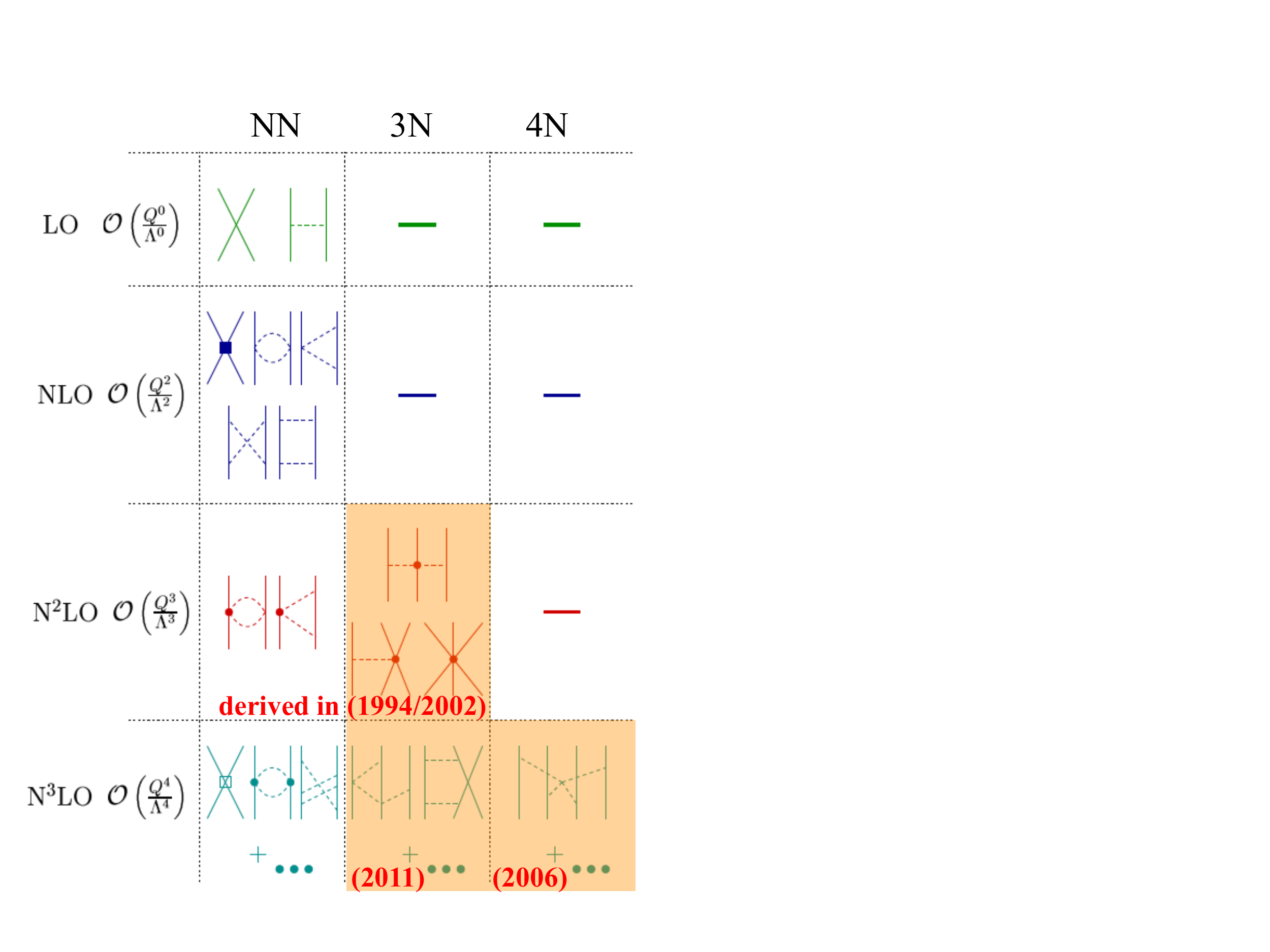}
\end{center}
\caption{Chiral EFT for nuclear forces, where the different contributions
at successive orders are shown diagrammatically [figure adapted
from~\cite{Epel09RMP,Mach11PR}].  Nucleons and pions are represented
by solid and dashed lines, respectively.  Many-body forces are
highlighted in orange including the year they were derived [3N forces
at N$^2$LO~\cite{Kolc94fewbody,Epel02fewbody} and
N$^3$LO~\cite{Bern083Nlong,Bern113Nshort}; 4N forces at
N$^3$LO~\cite{Epel064N}]. All N$^3$LO 3N and 4N forces are predicted
parameter-free.\label{fig:chiralEFT}}
\end{figure}

Chiral EFT provides a systematic basis for strong interactions at
momentum scales of the order of the pion mass $Q \sim m_\pi$ based on
the symmetries of quantum chromodynamics
(QCD)~\cite{Epel09RMP,Mach11PR}. In chiral EFT, nucleons interact via
pion exchanges and shorter-range contact interactions. The resulting
nuclear forces and consistent electroweak operators are organized in
a systematic expansion in powers of $Q/\Lambda_{\rm b}$, where
$\Lambda_{\rm b} \sim 500 \, {\rm MeV}$ denotes the breakdown scale,
leading to a typical expansion parameter $Q/\Lambda_{\rm b} \sim
1/3$. The EFT enables controlled calculations with theoretical error
estimates, which is especially important for exotic nuclei and
neutron-rich matter under extreme conditions in
astrophysics. Moreover, chiral EFT connects nuclear forces to the
underlying theory through lattice QCD~\cite{Bean10NPLQCD,Bric14JPG}.

Generally, nuclear forces are not observable and depend on a resolution scale
$\Lambda$, so that the nuclear Hamiltonian is given by
$H(\Lambda)=T(\Lambda)+V_{\rm NN}(\Lambda)+V_{\rm 3N}(\Lambda) +V_{\rm
4N}(\Lambda) \ldots$ As shown in \textbf{Figure~\ref{fig:chiralEFT}}, at a
given order, nuclear forces include contributions from one- or multi-pion
exchanges that govern the long- and intermediate-range parts and from short-
range contact interactions. For each $\Lambda$, the scale-dependent short-
range couplings are fit to low-energy data and thus capture all short-range
effects relevant at low energies. While 3N forces are not observable, there
are natural sizes to many-body-force contributions that are made manifest in
the EFT power counting and which explain the phenomenological hierarchy of
many-body forces, $V_{\rm NN}(\Lambda) > V_{\rm 3N}(\Lambda) > V_{\rm
4N}(\Lambda)$~\cite{Epel09RMP,Mach11PR}. The effects discussed in this review
are dominated by the long-range parts of 3N forces and are therefore expected
to be robust. Even though it is tempting to neglect contributions from 3N
forces in cases when calculations based on only NN forces provide a good
description of experimental data [see, e.g., Reference~\cite{Ekst13optNN}],
EFT power counting dictates the inclusion of all many-body forces up to a
given order. In fact, explicit calculations show that 3N forces always provide
important contributions in nuclei and matter~(see
Sections~\ref{oxygen},~\ref{calcium} and~\ref{matter}). The scale dependence
can also be exploited by using the renormalization group (RG) to
systematically change the resolution scale, while preserving low-energy
observables. This can be advantageous for calculations of nuclei and nucleonic
matter, because the evolution to lower scales facilitates the solution of the
nuclear many-body problem due to a decoupling of low and high momenta in the
Hamiltonian~\cite{Bogn09PPNP,Furn13RPP}.

\subsection{Many-body forces}
\label{mbforces}

Chiral EFT opens up a systematic path to investigate many-body forces
and their impact on few- and many-body systems~\cite{Hamm12RMP}. An
important feature of chiral EFT is the consistency of NN and 3N (and
higher-body) interactions. This determines the long-range
two-pion-exchange parts of 3N forces at next-to-next-to-leading order
(N$^2$LO), with pion-nucleon couplings $c_1, c_3, c_4$, leaving only
two low-energy couplings $c_D$ and $c_E$ that encode pion interactions
with short-range NN pairs and short-range three-body physics,
respectively~\cite{Kolc94fewbody,Epel02fewbody}. To fit $c_D$ and
$c_E$, different uncorrelated observables are used, e.g., the
binding energy and half life of $^3$H~\cite{Gazi08lec}, or the binding
energy of $^3$H and the charge radius of $^4$He~\cite{Hebe11fits}.

\begin{figure}[t]
\begin{center}
\includegraphics[width=\textwidth,clip=]{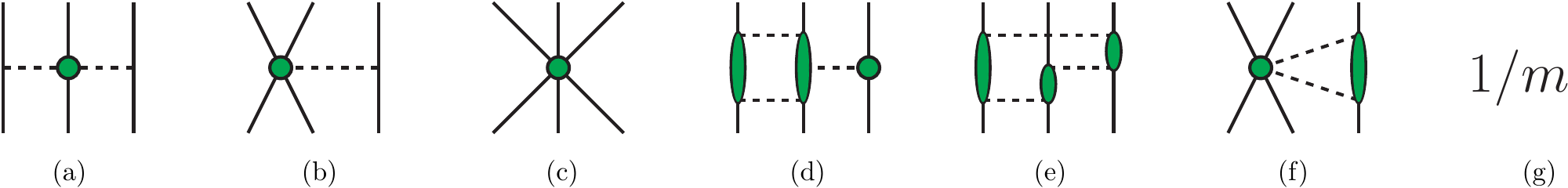}
\end{center}
\caption{Different topologies that contribute to 3N forces up 
to N$^3$LO. The shaded vertices denote the amplitudes of the
corresponding pion/nucleon interactions. The individual diagrams are:
(a)~2$\pi$ exchange, (b)~1$\pi$-contact, (c)~3N contact,
(d)~2$\pi$-1$\pi$ exchange, (e)~ring contributions,
(f)~2$\pi$-contact, and (g)~relativistic corrections. Figure taken
from Reference~\cite{Hebe15N3LOpw}. \label{fig:diag_N3LO}}
\end{figure}

At the next order, all many-body interactions are predicted
parameter-free with many new structures, as shown in
\textbf{Figure~\ref{fig:diag_N3LO}}. These also depend on the leading
NN contact interactions $C_S, C_T$~\cite{Bern083Nlong,Bern113Nshort}
[see the 2$\pi$-contact contributions (f) and the relativistic
corrections (g)]. Interestingly, for systems of only neutrons, the
N$^2$LO $c_D$ and $c_E$ parts do not contribute because of the Pauli
principle and the coupling of pions to
spin~\cite{Hebe10nmatt}. Therefore, chiral EFT predicts all
three-neutron and four-neutron forces to N$^3$LO.

The leading N$^2$LO 3N forces improve few-body scattering, but
interesting open problems remain~\cite{Nass12RPP}. This makes the
application of 3N and 4N forces at the next order (N$^3$LO) very
exciting. The derivation of N$^3$LO 3N forces has only been completed
recently~\cite{Bern083Nlong,Bern113Nshort}, but no calculation exists
for nuclei beyond $A=3$, where the current state-of-the-art are
N$^3$LO NN plus N$^2$LO 3N calculations, or consistent calculations at
N$^2$LO (see Sections~\ref{oxygen} and~\ref{calcium}). The practical
calculation of all topologies of Figure~\ref{fig:diag_N3LO} in a form
suitable for few- and many-body calculations is a nontrivial task. Due
to the large amounts of required computational resources, 3N matrix
elements so far have only been available in small basis
spaces~\cite{Gola14n3lo}. Recently, a novel and numerically much more
efficient method has been developed for decomposing 3N forces in a
plane-wave partial-wave basis~\cite{Hebe15N3LOpw}. The new framework
makes explicit use of the fact that all unregularized contributions to
chiral 3N forces are either local, i.e., they depend only on
momentum transfers, or they contain only polynomial non-local
terms. These new developments allow to calculate matrix elements of
all N$^3$LO 3N contributions for large basis spaces required for ab
initio studies of nuclei and nucleonic matter. An important open
problem is to fit all NN and 3N low-energy couplings at N$^3$LO
consistently. To this end, it may be beneficial to fit several
different few-body observables simultaneously within theoretical
uncertainties, or to include also information beyond few-nucleon
systems in the fits. The exploration of these strategies is currently
in progress.
\section{NEUTRON-RICH OXYGEN ISOTOPES}
\label{oxygen}

With a closed $Z=8$ proton shell, the oxygen isotopes provide an
exciting laboratory to study nuclear forces with a range of many-body
methods. The oxygen chain also exhibits remarkable aspects of exotic
nuclei. In oxygen, the location of the neutron dripline at neutron number
$N=16$ is anomalously close to the stable nuclei [at the same $N$ as
for carbon ($Z=6$) and nitrogen ($Z=7$)], see
Section~\ref{dripline}. Adding a single proton to oxygen is able to
bind six more neutrons to $^{31}$F. As discussed in
Section~\ref{Ospectra}, oxygen features two more shell closures in
$^{22}$O and $^{24}$O, where the latter doubly magic nucleus lies just
at the neutron dripline. In recent years, experimental advances have
even allowed explorations of oxygen isotopes beyond the dripline to
$^{26}$O, see Section~\ref{beyond}.

In the following, we discuss results for the neutron-rich oxygen
isotopes and neighboring nuclei based on chiral EFT interactions. In
particular, the softening of nuclear forces by RG
transformations~\cite{Bogn09PPNP,Furn13RPP} enabled many new
calculations of nuclei with a range of many-body methods.  The results
with 3N forces surveyed in this review have been mainly performed
using two different strategies for the Hamiltonian with different RG
approaches.  The first uses the RG to evolve NN interactions to
low-momentum interactions $\vlowk$~\cite{Bogn09PPNP}. At the NN level,
the results discussed here start from the $500\mev$ N$^3$LO potential
of Reference~\cite{Ente03EMN3LO}. Chiral 3N forces provide a general
low-momentum basis, so that the low-momentum NN interactions have been
combined with N$^2$LO 3N forces fit to the $^3$H binding energy and
the $^4$He charge radius~\cite{Hebe11fits}. We will refer to these
Hamiltonians as low-momentum NN+3N forces. Note that similar results
are obtained if the RG evolution at the NN level is replaced by a
similarity RG (SRG) evolution~\cite{Hebe11fits}. The second strategy
for the Hamiltonian uses a SRG transformation applied consistently to
chiral NN and 3N interactions~\cite{Jurg09SRG3N,Roth14SRG3N}. The
results discussed in this review start from the same NN potential and
include the same N$^2$LO 3N operators, but with local
regulators~\cite{Navr07local3N}, fit to the binding energy and half
life of $^3$H~\cite{Gazi08lec}. We will refer to these Hamiltonians as
SRG-evolved NN+3N-full Hamiltonian, or NN+3N-ind Hamiltonian if the
SRG evolution starts only from the NN potential, but induced (ind) 3N
interactions are kept. The latter corresponds to the results for the
``bare'' $500\mev$ N$^3$LO potential, if induced higher-body interactions
can be neglected.

Next, we briefly discuss the different ab initio many-body methods
used for medium-mass nuclei. These can be grouped into calculations
based on valence-space Hamiltonians and calculations obtained directly
in large many-body spaces. The latter treat all nucleons as active in
a large basis space and rely on different, controlled approximations
to solving the many-body problem of $A$ nucleons. For calculations
based on valence-space Hamiltonians, the number of active degrees of
freedom is reduced by treating the nucleus as a many-body system
comprised of a closed-shell core with the additional $A_v$ valence
nucleons occupying a truncated valence space. The valence-space
Hamiltonian, which is diagonalized exactly, includes configurations
from the large basis space via different many-body methods.

Great advances to access large basis spaces have been made with
coupled-cluster (CC) theory~\cite{Hage14rev}. The CC method starts
from a closed-shell reference state and includes correlations through
a similarity transformation $\overline{H} = e^{-T} H e^T$, where $H$
is the normal-ordered Hamiltonian. In state-of-the-art calculations
the cluster operator $T=T_1+T_2+\dots+T_A$, which generates
particle-hole excitations to all orders, is truncated at the singles
and doubles (CCSD) level, $T_1+T_2$, and includes triples excitations,
$T_3$, in a nonperturbative but approximate way. At the CCSD level,
$T_1$ and $T_2$ are obtained by solving the CC equations, which follow
from the reference state having no one-particle--one-hole or
two-particle--two-hole excitations. Equations-of-motion CC methods can
access ground and excited states of one- or two-particles-attached and
one-particle-removed systems from closed-shell nuclei.

A novel ab initio many-body method is the in-medium similarity
renormalization group (IM-SRG)~\cite{Tsuk10IMSRG,Herg13IMSRG}. The
IM-SRG uses a continuous unitary transformation $U(s)$, parameterized
by the flow parameter $s$, to drive the Hamiltonian to a band- or
block-diagonal form. This is accomplished by solving the flow equation
$\frac{dH(s)}{ds} =[\eta(s),H(s)]$, where $\eta(s) = \frac{dU(s)}{ds}
\, U^{\dagger}(s)$ is the generator of the transformation. With a
suitable choice of $\eta(s)$, the off-diagonal part of the Hamiltonian
is driven to zero as $s \rightarrow \infty$. Similarly to the CC
approach, the IM-SRG decouples the closed-shell ground state from the
space of particle-hole excitations on top of it. Recently, a
multi-reference formulation (MR-IM-SRG) was developed, which enables to
describe also ground states of open-shell nuclei~\cite{Herg13MR}.

In self-consistent Green's function theory (SCGF)~\cite{Barb09SCGF},
the quantity of interest is the single-particle Green's function,
which describes the propagation of single-particle and single-hole
excitations in the many-body system. From this the ground-state energy
can be calculated via the Koltun sum rule~\cite{Cipo13Ox}. The Gorkov
formalism~\cite{Soma11GGFform,Soma13GGF2N} allows to treat pairing
correlations explicitly and extends SCGF calculations to open-shell nuclei.

Other calculations of oxygen isotopes in large many-body spaces include
the importance-truncation no-core shell-model
(IT-NCSM)~\cite{Roth09ImTr,Roth11SRG}, which extends an exact
diagonalization with the NCSM by using importance sampling to access
larger spaces, and nuclear lattice simulations~\cite{Epel1416O}, which
solve the many-body problem of nucleons on a Euclidean space-time
lattice. In lattice EFT, the energies of the ground and excited states
are obtained by propagating the system in imaginary time, as in lattice QCD.

Approaches based on valence-space Hamiltonians allow the calculation
of ground and excited states of all nuclei in the valence space
provided that the diagonalization is
feasible~\cite{Caur05RMP}. Many-body perturbation theory
(MBPT)~\cite{Hjor95MBPT} has been used to derive valence-space
Hamiltonians, based on a diagrammatic approach for calculating the
single-particle energies (SPEs) and the interactions between valence
nucleons. The MBPT includes the contributions from configurations
outside the valence space perturbatively. State-of-the-art MBPT
results include NN and 3N forces to third order for the valence-space
interactions and the consistent SPEs [see Reference~\cite{Holt13Ox}
for oxygen isotopes].

Recently, nonperturabtive derivations of valence-space Hamiltonians
have been achieved based on the IM-SRG and CC methods. In the IM-SRG
for open-shell nuclei, states with $A_v$ particles in the valence
space are additionally decoupled from those containing non-valence
admixtures~\cite{Tsuk12SM,Bogn14SM}. This gives the energy of the
closed-shell core (as in the standard IM-SRG), but also valence-space
SPEs and interactions. In the CC effective interaction (CCEI)
approach, a similar decoupling for a given nucleus is achieved using a
Lee-Suzuki similarity transformation from the CC solution in the large
basis space to the valence space~\cite{Jans14SM,Ekst14GT2bc}.

\begin{figure}[t]
\begin{center}
\includegraphics[width=0.925\textwidth,clip=]{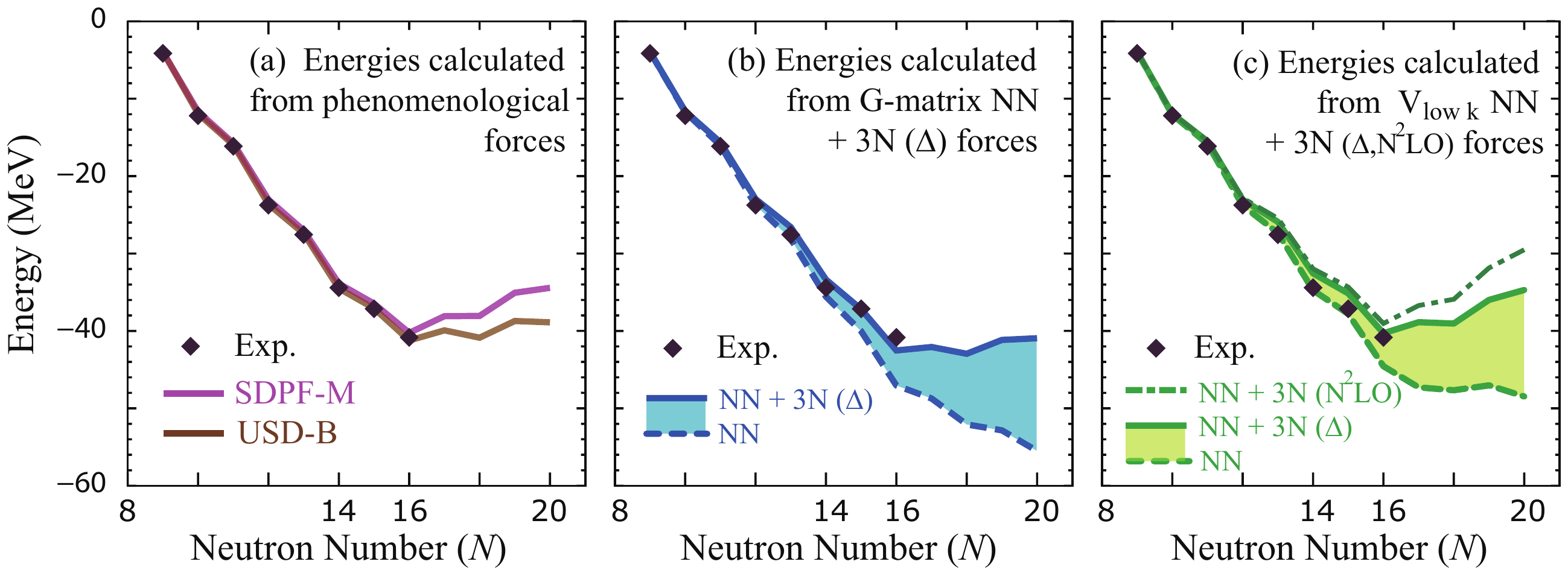} \\[2mm]
\minipage{0.45\textwidth}
\includegraphics[width=\linewidth,clip=]{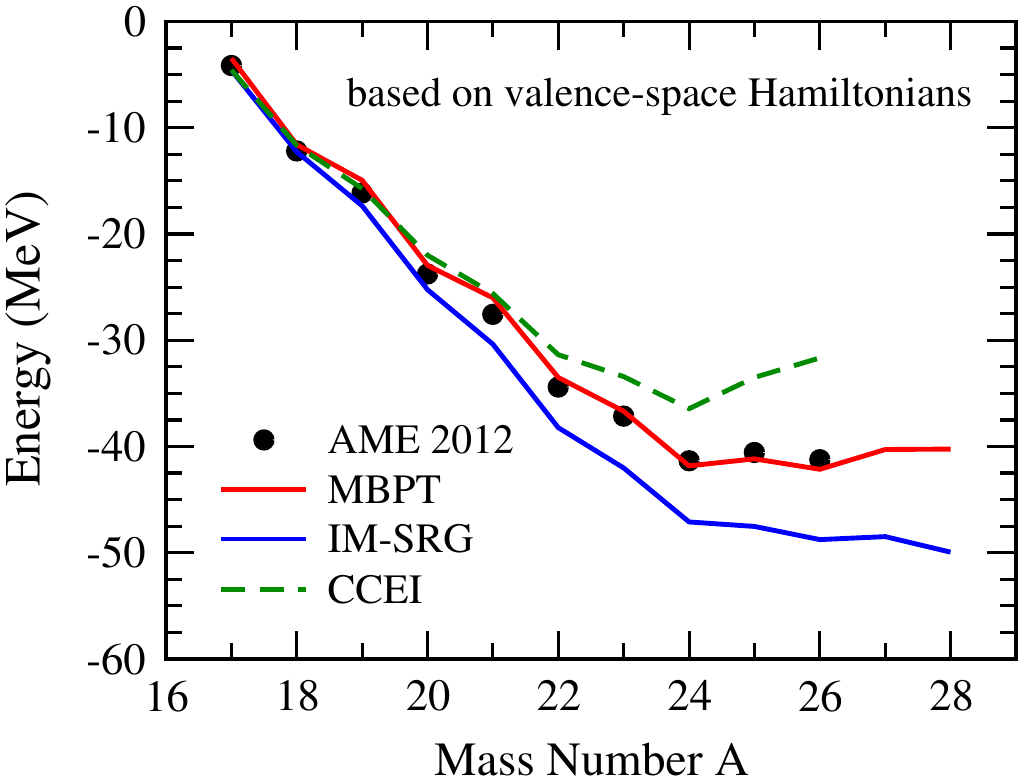}
\endminipage\hspace{2mm}
\minipage{0.45\textwidth}
\includegraphics[width=\linewidth,clip=]{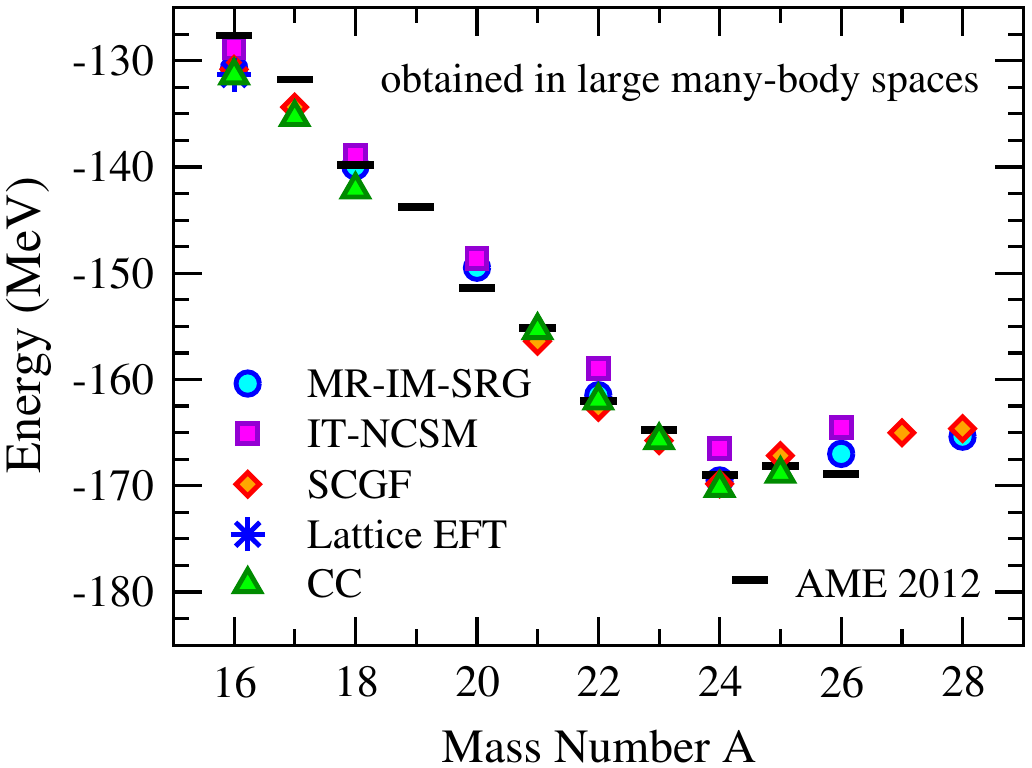}
\endminipage
\end{center}
\caption{Top panel:~Ground-state energies of oxygen isotopes measured 
from $^{16}$O, including experimental values of the bound $^{16-24}$O.
Figure taken from~\cite{Otsu10Ox}. Energies obtained from
(a)~phenomenological forces SDPF-M~\cite{Utsu04n20} and
USDB~\cite{Brow06USD}, (b)~a $G$ matrix and including Fujita-Miyazawa
3N forces due to $\Delta$ excitations, and (c)~from low-momentum
interactions $\vlowk$ and including N$^2$LO 3N forces as well as 
only due to $\Delta$ excitations. The changes due to 3N forces based
on $\Delta$ excitations are highlighted by the shaded areas.
Bottom~panels:~Left:~Ground-state energies of oxygen isotopes relative
to $^{16}$O based on valence-space Hamiltonians, compared to
the atomic mass evaluation (AME 2012)~\cite{Wang12AME12}. The MBPT
results are performed in an extended $sdf_{7/2}p_{3/2}$ valence
space~\cite{Holt13Ox} based on low-momentum NN+3N forces, while the
IM-SRG~\cite{Bogn14SM} and CCEI~\cite{Jans14SM} results are in the
$sd$ shell from a SRG-evolved NN+3N-full Hamiltonian. 
Right:~Ground-state energies obtained in large many-body
spaces: MR-IM-SRG~\cite{Herg13MR}, IT-NCSM~\cite{Herg13MR},
SCGF~\cite{Cipo13Ox}, CC~\cite{Jans14SM}, based on the SRG-evolved
NN+3N-full Hamiltonian, and Lattice EFT~\cite{Epel1416O}, based on
NN+3N forces at N$^2$LO.\label{fig:O}}
\end{figure}

\subsection{Location of the neutron dripline}
\label{dripline}

The neutron drip line evolves regularly from light to medium-mass
nuclei except for a striking anomaly in the oxygen isotopes, where the
dripline is at a doubly magic nucleus $^{24}$O and anomalously close
to the stable nuclei. This anomaly is challenging to explain in
microscopic theories based only on NN forces that reproduce NN
scattering~\cite{Otsu10Ox}. This is illustrated in the top panel of
\textbf{Figure~\ref{fig:O}} with $sd$-shell calculations based on
second-order MBPT [dashed lines in panels~(b) and (c)], where the
ground-state energies decrease up to $N=20$, leading to an incorrect
dripline at $^{28}$O. This is in contrast to phenomenological
interactions adjusted to experiment, shown in panel~(a), which have a
minimum at $^{24}$O because the $d_{3/2}$ orbital remains
unbound. The comparison shows that NN forces only lead to too
attractive interactions among valence neutrons, which causes the
$d_{3/2}$ orbital to become bound in $^{24}$O and beyond. These
deficiencies have been traced to the monopole components of the
valence-space Hamiltonian, as these parts are amplified with neutron
number, and it has been argued that 3N forces are important for the
monopole components~\cite{Zuke033N}. This is also supported by the
cutoff dependence of the monopole components with NN forces
only~\cite{Schw08mono}.

First investigations of neutron-rich oxygen isotopes with 3N
forces~\cite{Otsu10Ox} have shown that 3N forces lead to repulsive
interactions between valence neutrons. This is dominated by the
long-range parts of 3N forces, as highlighted by the results for
Fujita-Miyazawa 3N forces~\cite{Fuji573N} due to $\Delta$ excitations
in the top panel~(b) and (c) of Figure~\ref{fig:O}. This results from
the interactions of two valence neutrons with any of the nucleons in
the core, which corresponds to the normal-ordered two-body part of 3N
forces, and the repulsive nature for valence neutrons is rather
general~\cite{Otsu10Ox}. The dominance of the normal-ordered two-body
part, which is enhanced by all core nucleons, can be
understood based on phase-space arguments for normal Fermi
systems~\cite{Frim113Nres} and was verified in explicit
calculations~\cite{Hage07CC3N,Roth12NCSMCC3N}. The top panel~(c) of
Figure~\ref{fig:O} shows that N$^2$LO 3N forces, fit to few-nucleon systems
only, predict the dripline correctly. It is interesting to note that
the same 3N forces also lead to repulsion in neutron matter, see
Section~\ref{nmat}.

This repulsive 3N-force mechanism was confirmed in more recent
calculations based on large many-body spaces and with improved MBPT
and nonperturbative valence-space Hamiltonians, as shown in the bottom
panels of Figure~\ref{fig:O}. All results obtained in large many-body
spaces with a SRG-evolved NN+3N-full
Hamiltonian~\cite{Herg13MR,Cipo13Ox,Jans14SM} (see the bottom right
panel) lead to the correct dripline position at $^{24}$O and the
different many-body methods agree within a few percent. With NN+3N-ind
forces only (not shown), all oxygen isotopes are underbound with
respect to experiment and bound up to $^{28}$O. The results obtained
from ab initio valence-space Hamiltonians are shown in the bottom left
panel of Figure~\ref{fig:O}. The nonperturbative calculations are in
the $sd$ shell, while the MBPT calculations are performed in an
extended $sdf_{7/2}p_{3/2}$ valence space. In this case, the range of
predictions is broader, but the 3N contributions also lead to an
increased repulsion with neutron number. The broader range here is due
to the different Hamiltonians considered (MBPT vs.~IM-SRG/CCEI) as
well as due to different many-body approximations.

\begin{figure}[t]
\begin{center}
\minipage{0.335\textwidth}
\includegraphics[width=\linewidth,clip=]{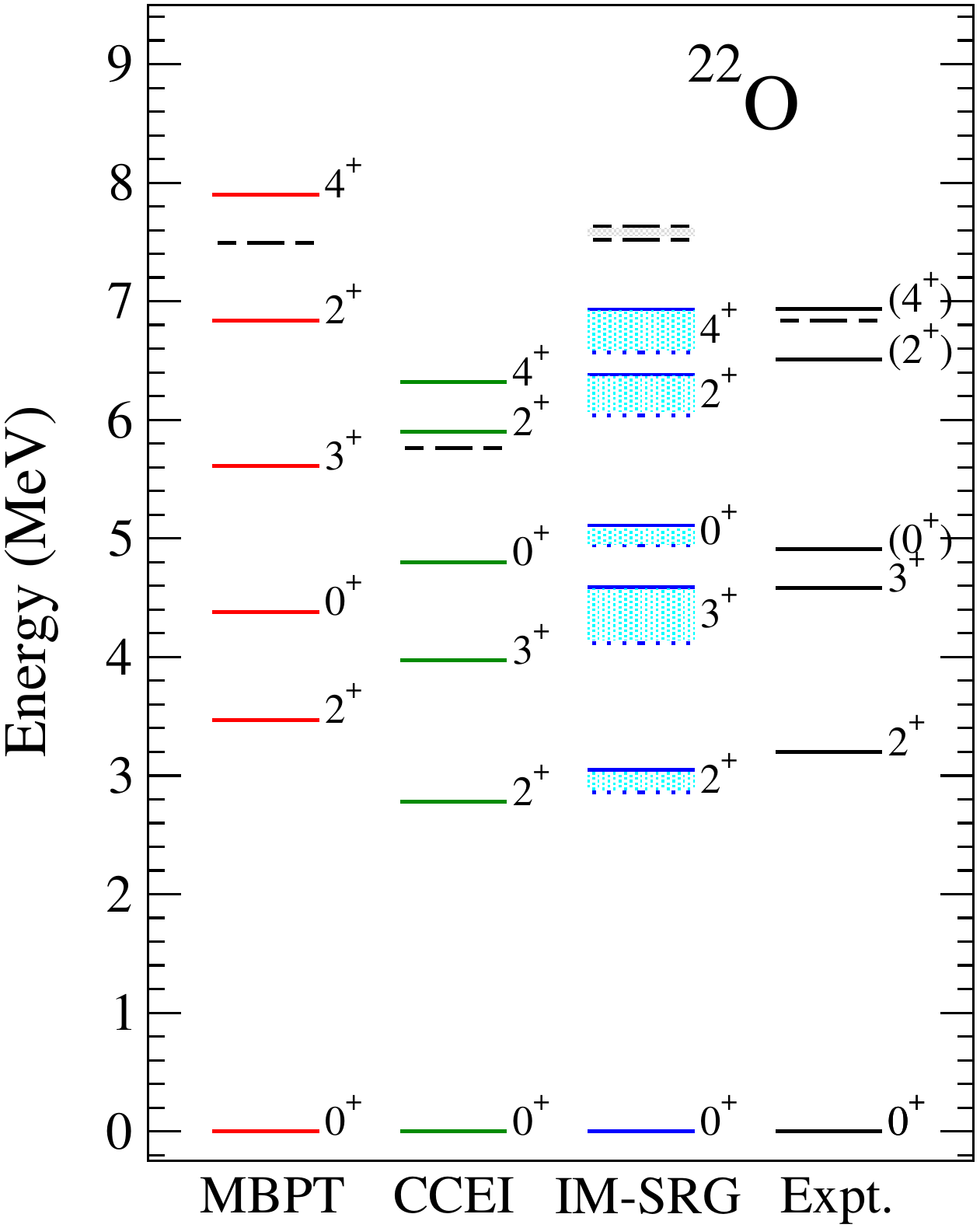}
\endminipage\hfill
\minipage{0.3\textwidth}
\includegraphics[width=\linewidth,clip=]{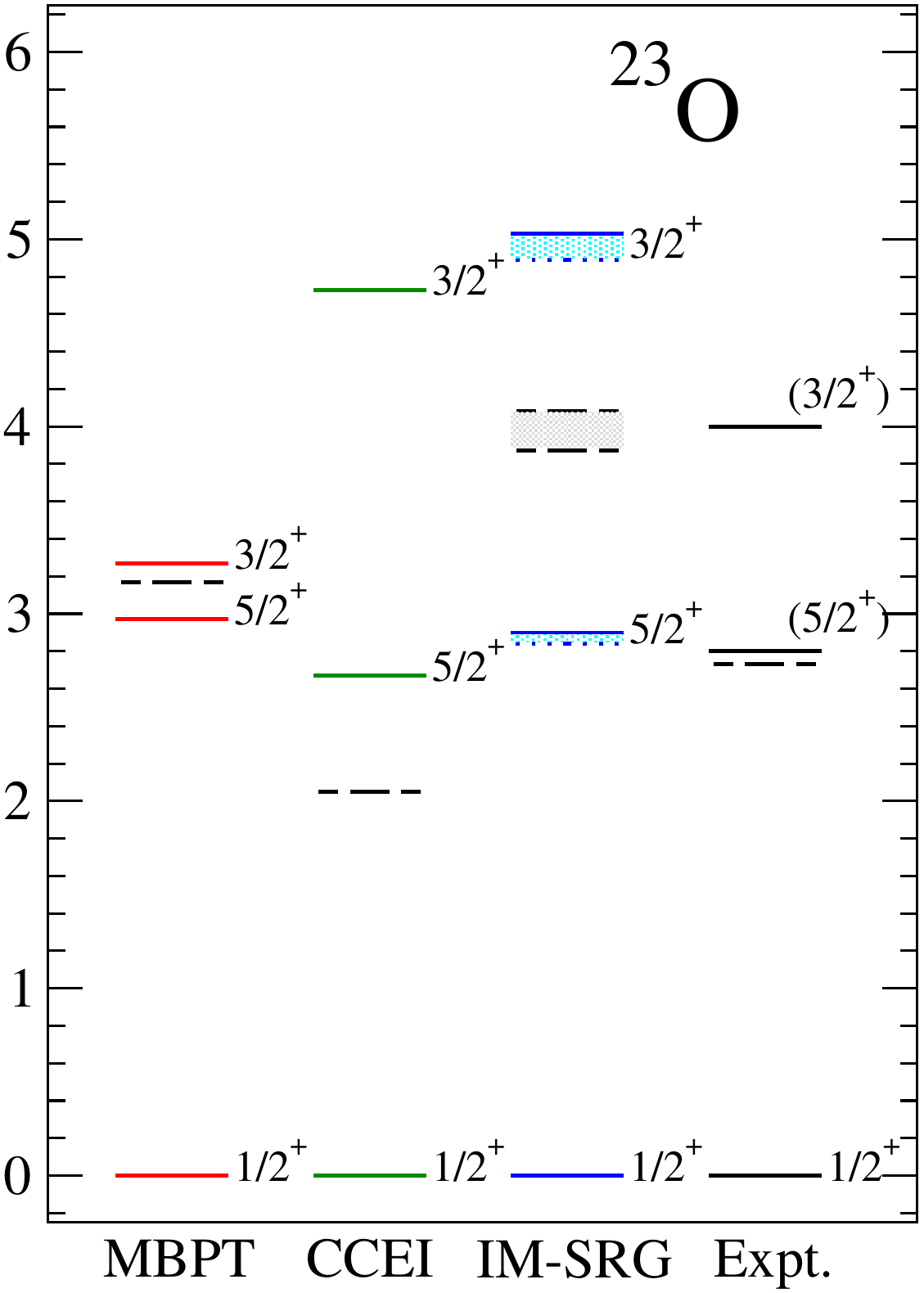}
\endminipage\hfill
\minipage{0.3\textwidth}
\includegraphics[width=\linewidth,clip=]{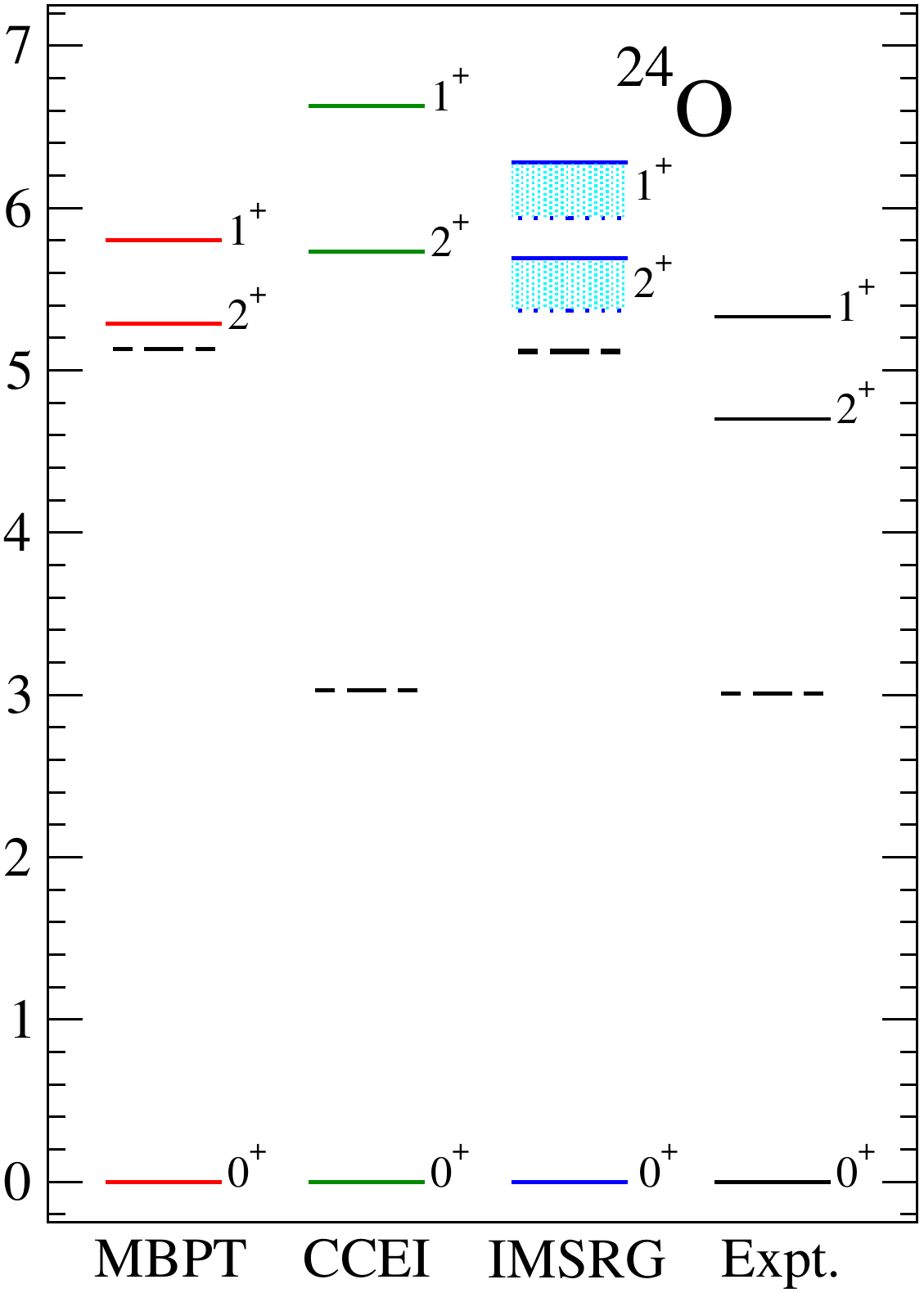}
\endminipage
\end{center}
\caption{Excited-state spectra of $^{22,23,24}$O based on NN+3N forces 
and compared with experiment. Figures adapted from~\cite{Bogn14SM}.
The MBPT results are performed in an extended $\sdfp$
space~\cite{Holt13Ox} based on low-momentum NN+3N interactions, while
the IM-SRG~\cite{Bogn14SM} and CCEI~\cite{Jans14SM} results are in the
$sd$ shell from the SRG-evolved NN+3N-full Hamiltonian with
$\hbar \omega=20 \mev$ (CCEI and dotted IM-SRG) and $\hbar
\omega=24\mev$ (solid IM-SRG). The dashed lines show the neutron
separation energy.\label{fig:Ospectra}}
\end{figure}

\subsection{Spectroscopy}
\label{Ospectra}

Next, we discuss excited states in the neutron-rich oxygen isotopes
$^{22,23,24}$O, which have been the subject of recent
experiments~\cite{Stan04Ox,Fern1122O,Elek0723O,Schi0723O,Hoff0924O,Tsho1224O}.
\textbf{Figure~\ref{fig:Ospectra}} compares the measured low-lying
states with theoretical calculations for each nucleus. We note that
the spectra of neutron-rich oxygen isotopes have also been calculated
in CC theory with phenomenological 3N forces adjusted to the oxygen
isotopes ($\tneff$) and including the continuum~\cite{Hage12Ox3N},
with good agreement to experiment. In Figure~\ref{fig:Ospectra}, we
show only the CCEI results, as they are based on the same SRG-evolved
interactions as in the IM-SRG calculations.

The first oxygen isotope with closed-shell properties for a
non-standard magic number, $^{22}$O, has its first $2^+$ state at
almost twice the energy as those in $^{18}$O and $^{20}$O. In contrast
to $^{24}$O, whose closed-shell nature can be qualitatively well
described with NN forces due to the large separation between the $\dt$
and $\so$ orbitals, the spectrum of $^{22}$O is not well reproduced
with NN forces: the first $2^+$ state is below experiment, and the
rest of the spectrum is generally too compressed.  Only when 3N forces
are included, the $2^+$ energy is in good agreement with
experiment. Results with NN+3N forces are shown in
Figure~\ref{fig:Ospectra} for MBPT calculations in an extended $\sdfp$
space~\cite{Holt13Ox} or with IM-SRG~\cite{Bogn14SM} and
CCEI~\cite{Jans14SM} in the $sd$ shell. For the next excited states,
the $2^+ - 3^+$ splitting is somewhat large in MBPT and results in an
inversion of the $3^+$ and $0^+$ states and a too-even spacing
of the other levels. This is not the case for IM-SRG and CCEI, where
the level ordering is well reproduced and the spacings between
states are close to experiment. Remarkably, the IM-SRG and CCEI
results are within less than $200 \kev$ when the same 
harmonic-oscillator value $\hw$ is used.

Of particular interest is the spectrum of $^{23}$O, which provides a
unique test for theory, as it simultaneously reflects the features of
the doubly magic $^{22}$O and $^{24}$O. The shell model $^{23}$O
ground state is dominated by one particle in the $\so$ orbit, while
the two lowest excited states are expected to be a single-particle
$5/2^+$ one-hole excitation, indicative of the strength of the
$^{22}$O shell closure, and a higher-lying single-particle $3/2^+$
one-particle excitation, reflecting the strength of the $^{24}$O shell
closure. The $5/2^+$ state lies just above the neutron decay
threshold, while the $3/2^+$ state resides in the
continuum. Reflecting the failure to reproduce the $^{22}$O shell
closure, all calculations with NN forces only predict a too low
$5/2^+$ state.  With 3N forces included, shown in the middle panel of
Figure~\ref{fig:Ospectra}, the $5/2^+$ state is well reproduced in all
calculations~\cite{Holt13Ox,Bogn14SM,Jans14SM}. The position of the
$3/2^+$ state is approximately $1.0 \mev$ too low in MBPT and
$1.0\mev$ too high with IM-SRG and CCEI, which are again in remarkable
agreement. The inclusion of the continuum is expected to lower the
$\dt$ orbital~\cite{Tsuk09cont} improving the IM-SRG and CCEI results.

Finally, the right panel of Figure~\ref{fig:Ospectra} shows the
spectrum of $^{24}$O with 3N forces in comparison with experiment.
All calculations find a clear closed-shell signature of a high $2^+$
energy. Moreover, the spacing to the next $1^+$ state is well
reproduced. Note that due to the unbound $\dt$ orbital, continuum
effects should be included~\cite{Hage12Ox3N}.

\subsection{Beyond the neutron dripline}
\label{beyond}

Nuclei beyond the neutron dripline play an important role in
understanding the behavior of extreme neutron-rich systems. The
unbound $^{25}$O and $^{26}$O isotopes are the current experimental
limits in oxygen~\cite{Hoff0824O,Lund1226O,Caes1326O}, and the
lifetime of $^{26}$O makes it the first candidate to exhibit
two-neutron radioactivity~\cite{Kohl1326O}. A crucial aspect neglected
in most of the calculations discussed here is the coupling to the
continuum, which has been shown to play an important role for the
physics of unbound states~\cite{Mich09GSM} and specifically for the
oxygen isotopes~\cite{Voly05contSM}. CC $\tneff$ calculations found a
typical contribution from continuum coupling to be on the order of
$200 \kev$ for the unbound states past $^{24}$O.

\begin{figure}[t]
\begin{center}
\includegraphics[width=0.675\textwidth,clip=]{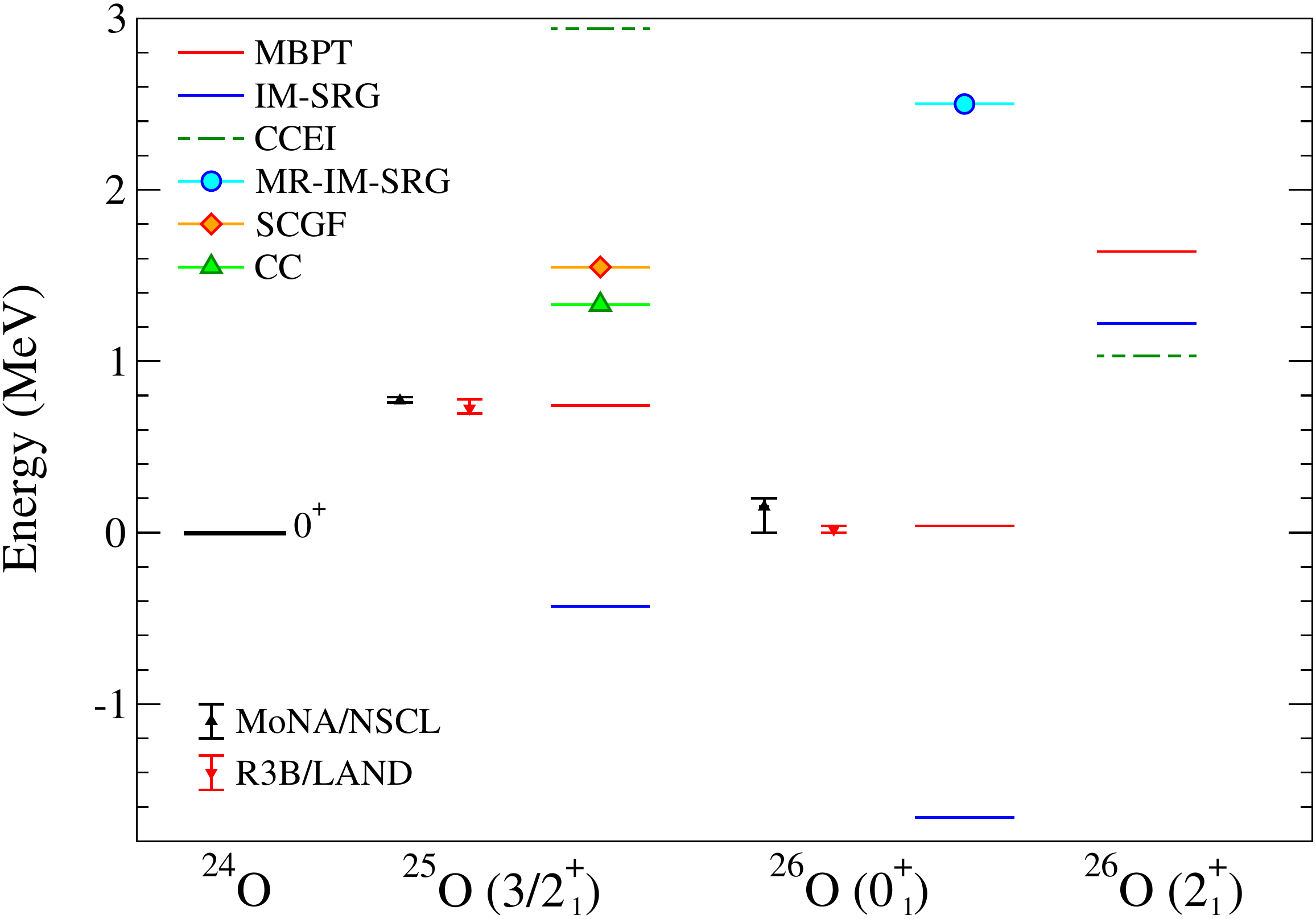}
\end{center}
\caption{Ground-state energies of $^{25}$O and $^{26}$O relative to
the $^{24}$O ground state, and the energy of the first excited state in
$^{26}$O relative to the $^{26}$O ground state. The experimental
energies are from MoNA/NSCL~\cite{Hoff0824O,Lund1226O} and from
R3B/LAND~\cite{Caes1326O}.  Results are shown for the different
many-body methods with NN+3N forces as in the bottom panels of
Figure~\ref{fig:O}: MBPT including also residual 3N
forces~\cite{Caes1326O}, IM-SRG~\cite{Bogn14SM}, CCEI~\cite{Jans14SM},
MR-IM-SRG~\cite{Herg13MR}, SCGF~\cite{Cipo13Ox}, and
CC~\cite{Jans14SM}.\label{fig:Odrip}}
\end{figure}

\textbf{Figure~\ref{fig:Odrip}} compares the experimental ground-state
energies of $^{25,26}$O to theoretical predictions with NN+3N forces.
In addition to the contribution from 3N forces to the SPEs and the
two-body interactions of valence neutrons, the MBPT results
shown~\cite{Caes1326O} also include the small contribution from
residual three-valence-neutron forces. These become more important
with increasing neutron number along isotopic
chains~\cite{Frim113Nres}, with a repulsive contribution from $0.1-0.4
\mev$ for $^{24-26}$O.  For the ground state of $^{25}$O,
Figure~\ref{fig:Odrip} shows that MBPT, CC, and SCGF agree well with
experiment, while the valence-space predictions from IM-SRG and CCEI
are modestly too bound and unbound, respectively.  For the ground
state of $^{26}$O only MBPT gives a result close to
experiment. Nevertheless the overbinding seen in IM-SRG, and the
underbinding obtained in MR-IM-SRG and CCEI (off the scale of the
plot) are not unreasonable given expected theoretical
uncertainties. Better agreement is found for the continuum shell
model~\cite{Voly05contSM} and in CC $\tneff$
calculations~\cite{Hage12Ox3N}.

As shown in Figure~\ref{fig:Odrip}, valence-space NN+3N calculations
consistently predict a low first excited $2^+$ state in $^{26}$O
between $1.0-1.6 \mev$. USDA and USDB interactions give a somewhat
higher energy of $1.9 \mev$ and $2.1 \mev$,
respectively~\cite{Brow06USD}. Experimentally, events have been seen
at $4 \mev$~\cite{Caes1326O}, but in calculations the next state
lies above $6 \mev$ due to the large $\dt - \so$ gap. Note that the
lowest excited states in $^{26}$O have been very recently measured
with high statistics at RIBF/RIKEN.

\subsection{Neighboring open-shell nuclei}
\label{openshell}

\begin{figure}[t]
\begin{center}
\minipage{0.55\textwidth}
\includegraphics[width=\textwidth,clip]{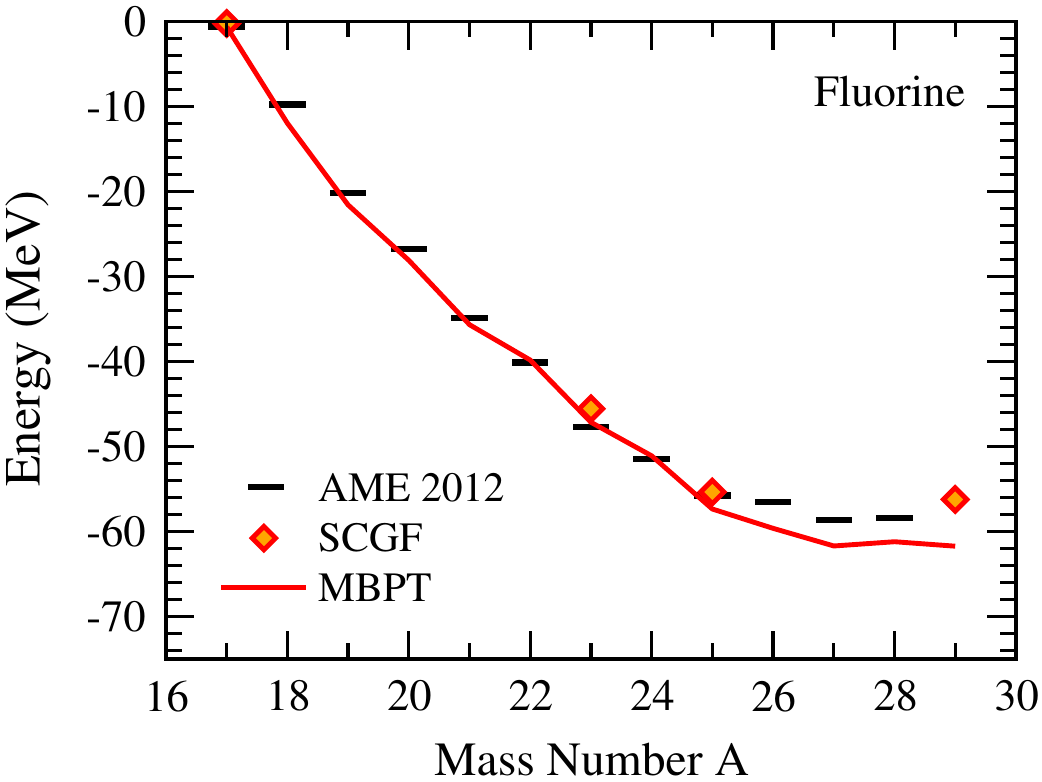}
\endminipage\hspace{5mm}
\minipage{0.335\textwidth}
\includegraphics[width=\linewidth,clip=]{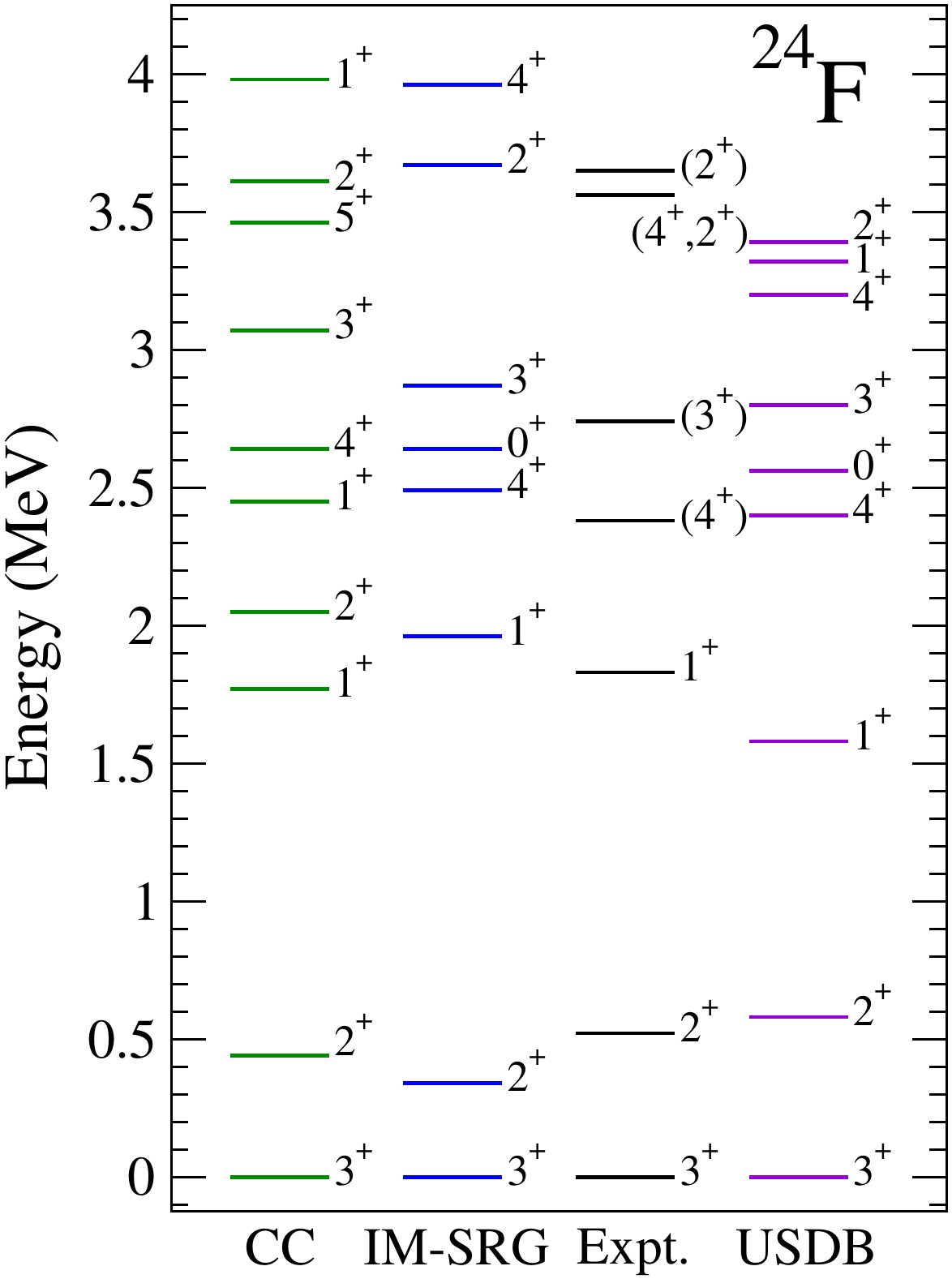}
\endminipage
\end{center}
\caption{Left panel:~Ground-state energies of fluorine isotopes 
measured from $^{16}$O, compared to AME 2012~\cite{Wang12AME12}.
The MBPT calculations~\cite{Simo15FNe} are performed in a proton $sd$
and neutron $\sdfp$ extended space based on low-momentum NN+3N forces,
while the SCGF results~\cite{Cipo13Ox} are with the SRG-evolved
NN+3N-full Hamiltonian. Right~panel:~Excited-state spectrum of
$^{24}$F compared with experiment~\cite{Cace1524F} and
USDB~\cite{Brow06USD}. The CC results~\cite{Ekst14GT2bc} are obtained
in a large many-body space based on the optimized chiral NN+3N forces
at N$^2$LO. The IM-SRG results~\cite{Bogn15FNe} are calculated in the
$sd$ shell from the SRG-evolved NN+3N-full Hamiltonian.\label{fig:F}}
\end{figure}

Fluorine isotopes have one more proton, and nuclear forces provide
more binding mainly due to the tensor part. No predictions based on
NN+3N forces exist for the dripline in fluorine, which lies at least
as far as $^{31}$F, but selected isotopes have been calculated from
SCGF~\cite{Cipo13Ox} and MBPT~\cite{Simo15FNe}. These are shown in the
left panel of \textbf{Figure~\ref{fig:F}}. Both SCGF and MBPT agree
well with experiment through $^{25}$F, with MBPT becoming modestly
overbound beyond. Future experiments are needed to test the
predictions beyond $^{28}$F.

Large-scale CC results for excited states have been presented for
$^{25,26}$F~\cite{Lepa1326F,Vajt1425F} with $\tneff$, and for
$^{22,24}$F with an optimized N$^2$LO NN+3N
interaction~\cite{Ekst14GT2bc}. In addition, the spectrum of $^{24}$F
has been studied recently with the IM-SRG in comparison to new
experimental results~\cite{Cace1524F}. These are shown in the right
panel of Figure~\ref{fig:F}, where we also compare to USDB results.
Without 3N forces (not shown), the spectrum is much too compressed and
the ordering of levels is incorrect for both CC and IM-SRG: The first
eight excited states lie below $2.0 \mev$, in clear contrast to
experiment. The NN+3N results for CC and IM-SRG shown in
Figure~\ref{fig:F} agree well with experiment, but differences are
seen due to the different starting Hamiltonians. For CC the predicted
first $2^+$ and $1^+$ states are well reproduced, but above these the
density and ordering of states begins to deviate from tentative
experimental spin-parity assignments.  While the ground-state energy
of $^{24}$F is overbound by $7.7 \mev$ in IM-SRG, the predicted
excited-state spectrum is in remarkably good agreement with the new
experimental measurements, with all excited states below the
one-neutron separation threshold less than $200 \kev$ away from
experiment.  The only exception is the $0^+$ state at $2.6 \mev$, also
predicted by USDB, to which the experiment is not sensitive.

\section{NEUTRON-RICH CALCIUM ISOTOPES}
\label{calcium}

The calcium isotopes ($Z=20$) provide an excellent region to study shell
evolution from stability towards the neutron dripline. In addition to
the standard doubly magic $^{40}$Ca and $^{48}$Ca, recent pioneering
experiments at rare isotope beam facilities have explored new shell
closures in exotic $^{52}$Ca and
$^{54}$Ca~\cite{Wien13Nat,Step13Ca54}. The wealth of spectroscopic
data on excited states and electromagnetic moments and transitions
offers excellent tests for nuclear forces.

The calcium and neighboring isotopic chains lie at the frontier of
theoretical calculations with 3N forces. Previous studies with
phenomenological valence-space interactions~\cite{Caur05RMP,%
Pove01KB3G,Honm05GXPF1A} or beyond-mean-field
calculations~\cite{Rodr07N32} provide a good description up to
$^{52}$Ca, but begin to disagree for more exotic isotopes where data
was not available. This provides an additional motivation for
calculations based on chiral EFT interactions.

The calcium isotopes $^{41-70}$Ca have been studied with MBPT in an
extended $\pfg$ valence space based on the same low-momentum NN+3N
forces as for
oxygen~\cite{Holt12Ca,Gall12Ca,Wien13Nat,Holt13pair,Holt14Ca}.  The
extended valence space compared to phenomenological interactions in
the $pf$ shell~\cite{Caur05RMP,Pove01KB3G,Honm05GXPF1A} suggests the
need to treat the $\gn$ orbital nonperturbatively in MBPT calculations
based on nuclear forces. Residual 3N forces between valence particles,
which are important for very neutron-rich nuclei as discussed in
Section~\ref{beyond}, provide very small contributions for the calcium
isotopes discussed here~\cite{Wien13Nat,Holt14Ca}. In addition to
valence-space calculations, the calcium isotopes have been studied in
large many-body spaces, where all nucleons are treated as active. The
neutron-rich calcium isotopes from $^{47-62}$Ca (except for
$^{57,58}$Ca) have been calculated in CC theory with phenomenological
$\tneff$ adjusted to the calcium
isotopes~\cite{Hage12Ca3N,Hage13Ca60}. In addition,
SCGF~\cite{Soma14GGF2N3N} and MR-IM-SRG~\cite{Herg14MR} provide
results for even calcium isotopes isotopes based on the same
SRG-evolved NN+3N-full Hamiltonian as for the oxygen
isotopes. Finally, based on the same Hamiltonian, there are IT-NCSM
results for the doubly magic $^{40,48}$Ca~\cite{Roth12NCSMCC3N}.

The MBPT and CC results for the ground-state energies are in very good
agreement with
experiment~\cite{Gall12Ca,Wien13Nat,Holt14Ca,Hage12Ca3N}, while the
SCGF~\cite{Soma14GGF2N3N} and MR-IM-SRG~\cite{Herg14MR} both give
overbound calcium isotopes, suggesting that the SRG-evolved NN+3N-full
Hamiltonian is too attractive for heavier nuclei. Nevertheless this
overbinding is systematic for all isotopes, and the experimental trend
is reasonably reproduced.

\subsection{Shell structure}
\label{shells}

\begin{figure}[t]
\begin{center}
\includegraphics[width=0.6\textwidth,clip=]{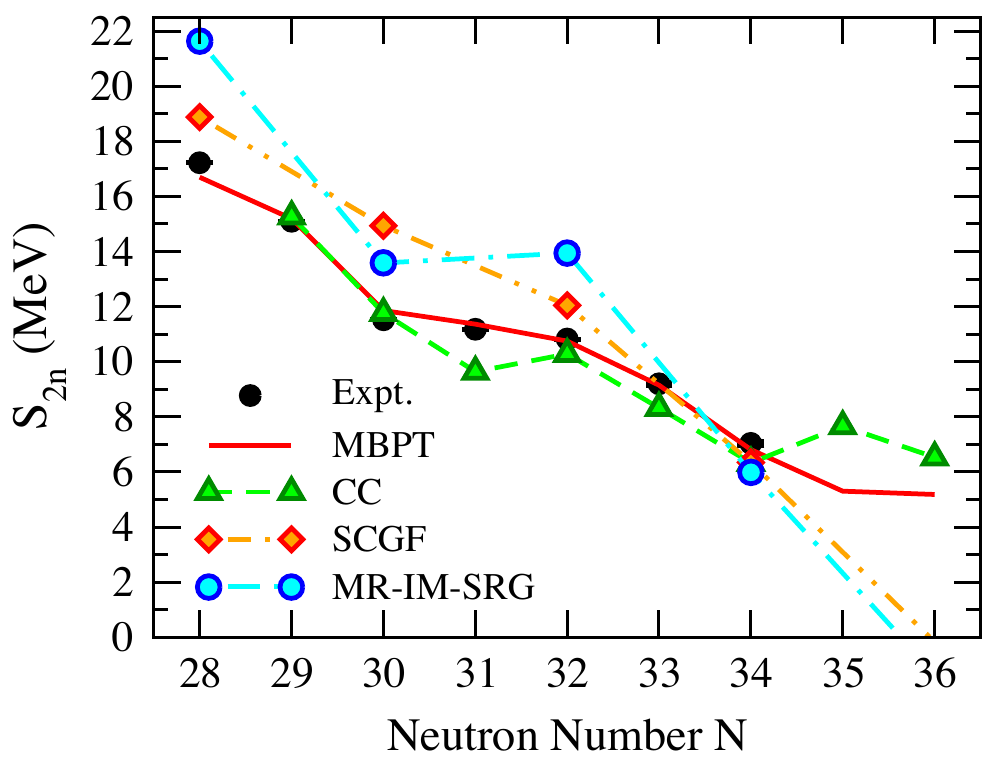}
\end{center}
\caption{Two-neutron separation energy $S_{2n}$ in neutron-rich calcium 
isotopes as a function of neutron number $N$. The experimental
energies~\cite{Wang12AME12,Gall12Ca,Wien13Nat} are compared with MBPT
predictions~\cite{Gall12Ca,Wien13Nat} based on low-momentum NN+3N
forces and CC theory with phenomenological $\tneff$~\cite{Hage12Ca3N}.
In addition, SCGF~\cite{Soma14GGF2N3N} and MR-IM-SRG~\cite{Herg14MR} 
results are shown based on a SRG-evolved NN+3N-full
Hamiltonian.\label{fig:Ca_S2n}}
\end{figure}

The shell evolution along an isotopic chain can be studied with the
two-neutron separation energy $S_{2n}$, where a significant decrease
occurs past a shell closure. \textbf{Figure~\ref{fig:Ca_S2n}} compares
the experimental $S_{2n}$ with calculated
MBPT~\cite{Gall12Ca,Wien13Nat}, CC~\cite{Hage12Ca3N},
SCGF~\cite{Soma14GGF2N3N} and MR-IM-SRG~\cite{Herg14MR} results. A
first key feature in Figure~\ref{fig:Ca_S2n} is the large decrease in
$S_{2n}$ from $N=28$ to $N=30$, a signature of the $N=28$ magic
number. This shell closure is not reproduced with NN forces
only~\cite{Holt12Ca,Hage12Ca3N,Caur05RMP}. First calculations with 3N
forces~\cite{Holt12Ca,Hage12Ca3N} showed that 3N forces are essential
for the $N=28$ shell closure. This also holds for the improved MBPT
calculations of References~\cite{Gall12Ca,Wien13Nat}, shown in
Figure~\ref{fig:Ca_S2n}, where 3N forces are included to third order
in MBPT [3N forces were only included to first order
in~\cite{Holt12Ca}]. While the $^{48}$Ca $S_{2n}$ has not been
calculated in CC, the decrease from $^{49}$Ca to $^{50}$Ca reproduces
experiment very well. Also shown in Figure~\ref{fig:Ca_S2n} are
SCGF~\cite{Soma14GGF2N3N} and MR-IM-SRG~\cite{Herg14MR} results, based
on the same SRG-evolved NN+3N-full Hamiltonian, which somewhat
underestimate (SCGF) or overestimate (MR-IM-SRG) the decrease in
$S_{2n}$ past $N=28$.

The flat behavior in $S_{2n}$ from $N=30$ to $N=32$ was predicted in
calculations with 3N forces~\cite{Holt12Ca,Gall12Ca}, as well as with
phenomenological shell-model
interactions~\cite{Pove01KB3G,Honm05GXPF1A}.  This was recently
confirmed with precision Penning-trap mass measurements of
$^{51,52}$Ca at TITAN/TRIUMF~\cite{Gall12Ca}, which found $^{52}$Ca to
be $1.74 \mev$ more bound. This represented the largest change in the
atomic mass evaluation in the last ten years. More recently, the
masses of $^{53,54}$Ca were measured in pioneering multi-reflection
time-of-flight mass measurement at ISOLTRAP/CERN~\cite{Wien13Nat}. The
resulting $S_{2n}$ values show a decrease from $N=32$ to $N=34$,
similar to the one past $^{48}$Ca. This unambiguously establishes the
doubly magic character of $^{52}$Ca. The shell closure in calcium at
$N=32$ had already been suggested based on the first excited $2^+$
energy in $^{52}$Ca~\cite{Huck85N32,Gade06Ca52} and nuclear
spectroscopy~\cite{Mant08Cabeta,Craw10Cabeta}. Figure~\ref{fig:Ca_S2n}
shows that the experimental $S_{2n}$ from $N=32$ to $N=34$ are in
excellent agreement with the MBPT and CC predictions. We also observe
that the different calculations with 3N forces start to deviate most
in $^{56}$Ca. Therefore, a future mass measurement of $^{56}$Ca would
provide a key test for theory, as well as direct information about the
closed-shell nature of $^{54}$Ca. Phenomenological shell-model
interactions~\cite{Pove01KB3G,Honm05GXPF1A} also provide a good
description of $S_{2n}$ to $^{54}$Ca, while energy-density
functionals~\cite{Erle12Nature} tend to predict an almost linear trend
in $S_{2n}$, missing the steeper decrease at $N=28$ and
$N=32$~\cite{Wien13Nat}.

\begin{figure}[t]
\begin{center}
\includegraphics[width=0.6\textwidth,clip=]{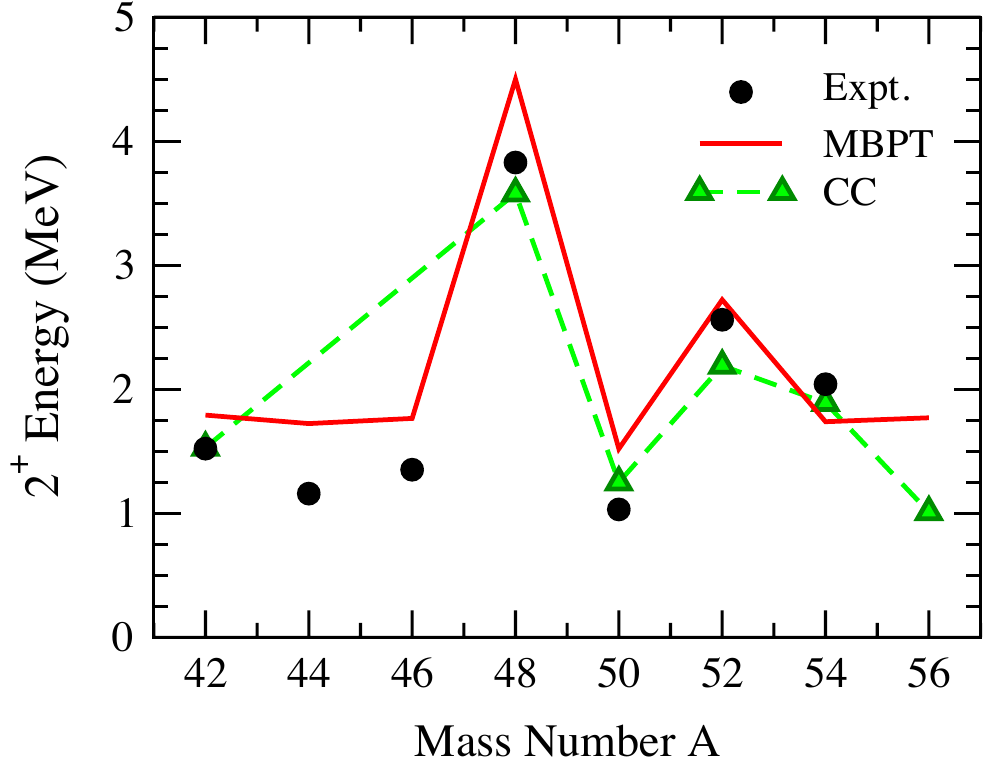}
\end{center}
\caption{Excitation energy of the first $2^+$ state in the even calcium
isotopes as a function of mass number~$A$. The MBPT~\cite{Holt14Ca}
and CC results~\cite{Hage12Ca3N} corresponding to the $S_{2n}$
calculations of Figure~\ref{fig:Ca_S2n} are compared to experiment
from~\cite{Step13Ca54,nndc14ENSDF}.\label{fig:Ca_2plus}}
\end{figure}

Another key observable to assess shell evolution is the excitation
energy of the first $2^+$ state in the even
isotopes. \textbf{Figure~\ref{fig:Ca_2plus}} shows the experimental
energies compared to the available theoretical calculations with NN+3N
forces:~MBPT \cite{Holt13pair} and CC \cite{Hage12Ca3N}. The agreement
of both calculations to experiment is very good, especially in
capturing the high $2^+$ energy in $^{48}$Ca and $^{52}$Ca, associated
with the $N=28$ and $N=32$ shell closures.  Especially interesting is
the $2^+$ energy in the exotic $^{54}$Ca, recently measured at
RIBF/RIKEN~\cite{Step13Ca54}, which is also in very good agreement
with both MBPT and CC predictions. The relatively high energy of this
$2^+$ state, and the increase compared to other $N=34$ isotones,
suggests a shell closure in calcium at $N=34$~\cite{Step13Ca54}.

From Figures~\ref{fig:Ca_S2n} and~\ref{fig:Ca_2plus}, we see that the
predictions of shell evolution are consistent when considering
different observables. In addition, three-point mass differences,
defined as $\Delta_n^{(3)} = \frac{(-1)^N}{2}[B(N+1,Z)+B(N-1,Z)
-2B(N,Z)]$ with (negative) ground-state energy $B(N,Z)$, provide
another signature for shell evolution, with relative peaks in
$\Delta_n^{(3)}$ associated with closed shells. When comparing MBPT
results~\cite{Holt13pair} for $\Delta_n^{(3)}$ with experiment, the
trend with neutron number along the calcium chain is well reproduced,
with peaks at $N=28$ and $N=32$, and an anomalously low
$\Delta_n^{(3)}$ value for $^{53}$Ca, placed between two shell
closures and dominated by the low-$j$ $p_{1/2}$ orbital~\cite{Brow13Gap}.

With present calculations based on NN+3N forces, it is difficult to
predict the neutron dripline because of the very flat behavior of the
ground-state energies past $^{60}$Ca found in energy-density
functional calculations, MBPT, CC, and
MR-IM-SRG~\cite{Hage12Ca3N,Holt14Ca,Herg14MR,Fors13CaEDF}, so that
very small interaction contributions can be decisive. In addition, for
such weakly bound systems, the continuum, currently included only in
CC, plays an important role~\cite{Hage12Ca3N}, and should also be
included in all calculations. Moreover, valence-space Hamiltonian for
more exotic isotopes will have to consider the $1d_{5/2}$ and
$2s_{1/2}$ orbitals~\cite{Hage12Ca3N,Lenz10LNPS}. Improved
calculations towards the dripline will allow the exploration of a
possible $N=40$ shell closure in $^{60}$Ca, which is the target of
ongoing experimental efforts in neighboring
isotopes~\cite{Gade14Ti60}. Finally, CC calculations combined with
halo EFT have suggested interesting Efimov physics around $^{60}$Ca,
with the possibility of $^{62}$Ca being a two-neutron halo
nucleus~\cite{Hage13Ca60}.

In addition to the calcium results, SCGF calculations have been
performed for the ground-state energies of Ar ($Z=18$), K ($Z=19$),
Sc ($Z=21$), and Ti ($Z=22$)~\cite{Soma14GGF2N3N},
and the resulting $S_{2n}$ generally agree
well with experiment, apart from an overestimated shell gap at $N=20$
attributed to the NN+3N-full Hamiltonian used. In $^{37-49}$K, the
evolution of the splitting between the lowest $1/2^+$ and $3/2^+$
states calculated with SCGF also successfully describes
experiment~\cite{Papu14Kmoments}.

Finally, 3N forces provide similar repulsive contributions to the
ground-state energies of the proton-rich $N=20$ isotones, which have
been studied in MBPT including the isospin-symmetry-breaking parts of
chiral EFT interactions~\cite{Holt13PR}. Ground-state properties of
heavier neutron-rich systems have also been explored in large-scale
calculations with NN+3N forces. First MR-IM-SRG results along the
nickel chain ($Z=28$)~\cite{Herg14MR}, based on the same NN+3N-full Hamiltonian
used in calcium, show a similar overbinding, also seen in CC
calculations for doubly magic nuclei up to
$^{132}$Sn~\cite{Bind14CCheavy}.

\subsection{Spectroscopy}
\label{Caspectra}

The study of shell evolution is complemented by exploring
spectroscopic properties of calcium isotopes based on NN+3N
forces. Spectra beyond $2^+$ states based on chiral NN+3N forces have
been obtained with MBPT~\cite{Holt14Ca} and with CC $\tneff$ for
states dominated by ground-state particle-hole excitations in
$^{52-56}$Ca~\cite{Hage12Ca3N}. Electromagnetic moments and
transitions have also been calculated with
MBPT~\cite{Holt12Ca,Holt14Ca,Garc15Camoment}.

The overall agreement of MBPT and CC spectra with experiment is good,
with low-lying states in odd-mass nuclei consistent with the shell
evolution discussed in Section~\ref{shells}. Some deficiencies are
present, e.g., with too low $5/2^-$ states in $^{49,51}$Ca in MBPT,
but generally the results are similar to phenomenological
interactions~\cite{Holt14Ca}. Low-lying excited states in $^{42}$Ti,
calculated with MBPT~\cite{Holt13PR}, and in neutron-rich
$^{52,54,56}$Ti, obtained from CC~\cite{Hage12Ca3N}, are also in good
agreement with experiment. Future experiments with rare isotopes can
test MBPT and CC predictions for several excited states in
neutron-rich $^{52-56}$Ca. In addition, there are MBPT predictions for
the mostly unexplored excited states in proton-rich $N=20$ isotones up
to $^{48}$Ni~\cite{Holt13PR}.

\begin{figure}[t]
\begin{center}
\includegraphics[width=0.65\textwidth,clip=]{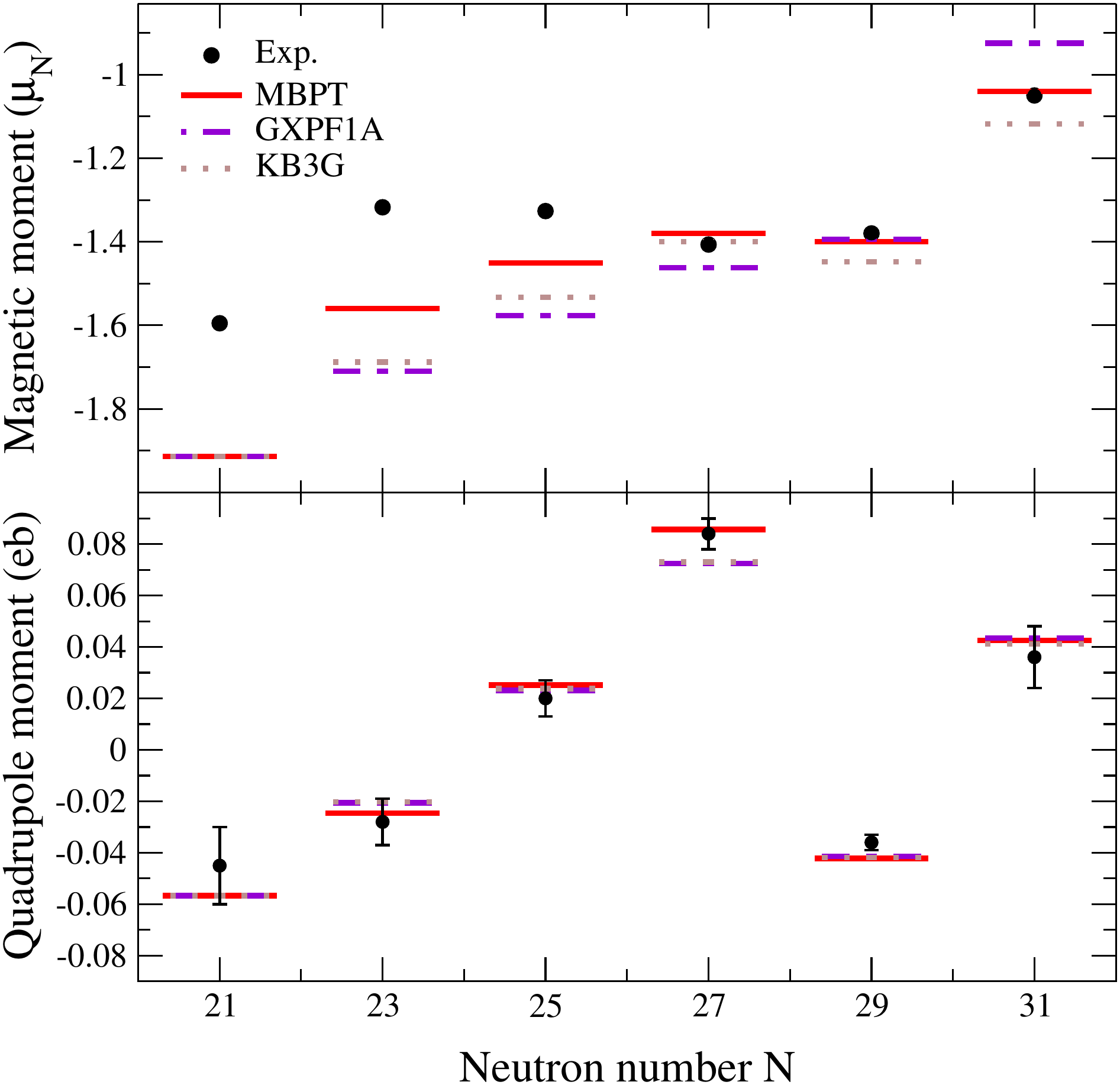}
\end{center}
\caption{Magnetic moments (upper panel) and electric quadrupole moments
(lower panel) of the odd calcium isotopes as a function of neutron
number $N$. Experiment~\cite{Garc15Camoment} is compared to MBPT
calculations with low-momentum NN+3N forces in the $\pfg$ valence
space~\cite{Garc15Camoment}. For comparison, $pf$-shell calculations
with phenomenological interactions KB3G~\cite{Pove01KB3G} and
GXPF1A~\cite{Honm05GXPF1A} are shown. Figure adapted 
from~\cite{Garc15Camoment}.\label{fig:Ca_moments}}
\end{figure}

Electromagnetic moments provide a complementary test of nuclear
forces. \textbf{Figure~\ref{fig:Ca_moments}} compares the experimental
magnetic moments and electric quadrupole moments of neutron-rich
calcium isotopes with MBPT predictions, including the very recent
measurements at COLLAPS/ISOLDE~\cite{Garc15Camoment}. For comparison,
the results with KB3G and GXPF1A interactions are also shown. The
lower panel of Figure~\ref{fig:Ca_moments} shows the electric
quadrupole moments. The experimental linear trend from $^{41}$Ca to
$^{47}$Ca and $^{49}$Ca to $^{51}$Ca, a signature of the filling of
the $f_{7/2}$ and $p_{3/2}$ orbitals, suggests that the quadrupole
moments are dominated by the single-particle character of the ground
states. The MBPT predictions exhibit a very good description of the
experimental quadrupole moments, in general similar to KB3G and GXPF1A
interactions and better for $^{47}$Ca. Note that the theoretical
results use the same neutron effective charges $e_n=0.5e$~\cite{Garc15Camoment}.
Electric quadrupole transitions involving excited states in
$^{46-50}$Ca have also been calculated in MBPT with the same effective
charges~\cite{Holt14Ca}. Similar to phenomenology, the agreement with
experiment is reasonable, taking into account that the measured
$B(E2)$ values vary within a factor of 50.

The upper panel of Figure~\ref{fig:Ca_moments} compares the
experimental magnetic moments with theoretical results obtained with
bare $g$-factors.  Calculations for the lighter isotopes
$^{41,43,45}$Ca, which assume a $^{40}$Ca core, do not reproduce
experiment, suggesting the importance of $sd$-shell degrees of freedom
for magnetic moments. This is in agreement with the $g$-factors of the
$2^+$ states in $^{42,44,46}$Ca~\cite{Schi03Ca44gfac,Tayl05Ca46gfac},
as well as calcium isotope shifts~\cite{Caur01Cashifts}. For $^{47}$Ca
MBPT reproduces experiment well, and the predictions for neutron-rich
$^{49}$Ca and $^{51}$Ca are in very good agreement with very recent
measurements~\cite{Garc15Camoment}. Magnetic moments with
phenomenological interactions are similar to MBPT except for
$^{51}$Ca, where the NN+3N calculation lies between KB3G and GXPF1A.

The magnetic dipole $B(M1)$ strength in $^{48}$Ca was first calculated
with MBPT in Reference~\cite{Holt12Ca}, with improvements in
Reference~\cite{Holt14Ca} by including 3N forces to third order. The
experimental peak at $10.2 \mev$~\cite{Neum98Ca48M1} is very well
reproduced in the MBPT calculation, but with too much strength,
suggesting a modest quenching of $q=0.9$ in the spin
$g$-factor. Phenomenological interactions GXPF1A and KB3 also
reproduce the concentrated peak and strength, but with a more
substantial quenching of $q=0.75$~\cite{Neum98Ca48M1}.

This quenching is in contrast with the calculations of the magnetic
moments in Figure~\ref{fig:Ca_moments}, where the bare spin
$g$-factors give a good description of experiment. This inconsistency
and the sensitivities to different interactions in
Figure~\ref{fig:Ca_moments} (compared to the smaller spread for the
quadrupole moments) point to the need for systematic calculations of
magnetic moment operators in the valence space. Necessary improvements
are the inclusion of electromagnetic two-body currents (or
meson-exchange currents), which are derived in chiral EFT consistently
with nuclear forces, as well as controlled calculations of effective
operators. Results with chiral two-body currents in light nuclei
demonstrate that they provide significant contributions to magnetic
moments~\cite{Past13momM12b}, while first applications to medium-mass
nuclei have focused on Gamow-Teller
transitions~\cite{Ekst14GT2bc,Mene110nbb2bc}.

\section{NEUTRON-RICH MATTER AND NEUTRON STARS}
\label{matter}

The physics of neutron-rich matter covers a wide range of extremes. At
very low densities, the average interparticle distance is sufficiently
large so that details of nuclear forces are not resolved and all
properties of the system are governed by the large S-wave scattering
length. In this universal regime, neutron matter shares many
properties with cold atomic gases close to the unitary limit, which
are the subject of active experimental and theoretical
studies~\cite{Zwer12book}. At intermediate densities, which are most
relevant for nuclei, the properties of neutron and nuclear matter are
used to guide the development of energy density functionals, in
particular to constrain the physics of neutron-rich systems, which are
key for understanding the synthesis of heavy nuclei in the
universe. At higher densities, far beyond nuclear densities, the
composition and properties of nuclear matter are still unknown. Exotic
states of matter containing strange particles or quark matter may be
present. On the other hand, neutron matter constitutes also a unique
laboratory for chiral EFT, because all many-body forces are predicted
to N$^3$LO, see Section~\ref{mbforces}. This offers the possibility to
derive reliable constraints based on chiral EFT interactions for the
equation of state (EOS) of neutron-rich matter in astrophysics, for
the symmetry energy and its density dependence, and for the structure
of neutron stars, but also makes it possible to test the chiral EFT
power counting and the hierarchy of many-body forces at densities
relevant for nuclei.

The importance of chiral 3N forces for understanding and predicting
nuclei has already been discussed in Sections~\ref{oxygen}
and~\ref{calcium}. The same 3N forces play an important role for
nuclear matter. In particular, the saturation of symmetric nuclear
matter is driven by 3N forces~\cite{Bogn05nuclmat,Hebe11fits}. While
3N contributions to neutron matter are smaller, they are crucial for
the EOS of neutron-rich matter, and thus for the symmetry energy and
its density dependence discussed in Section~\ref{symen}, and for
neutron stars in Section~\ref{nstar}.

\subsection{Neutron matter properties and theoretical uncertainties}
\label{nmat}

\begin{figure}[t]
\begin{center}
{\minipage{0.495\textwidth}
\includegraphics[width=\linewidth,clip=]{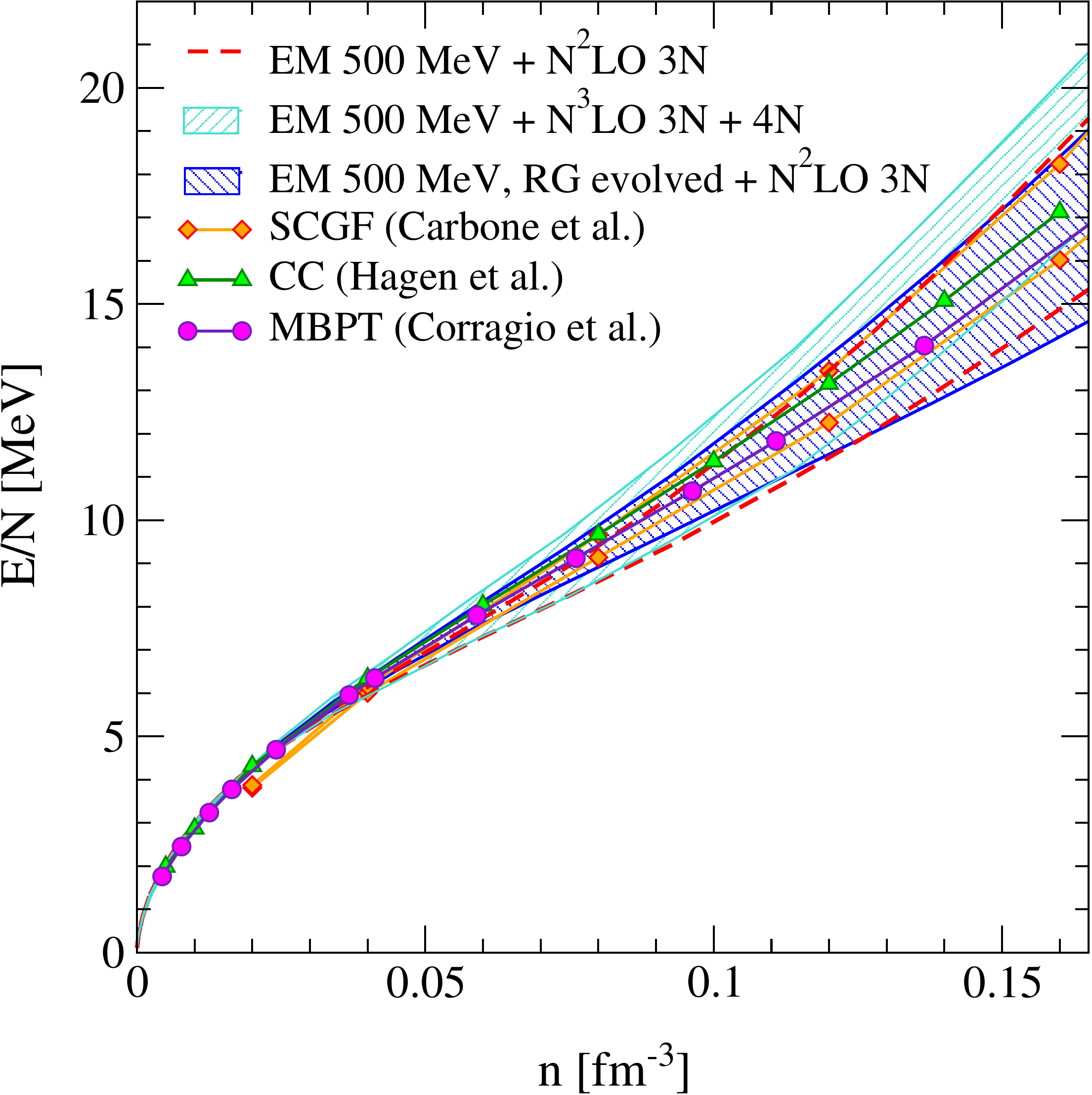}
\endminipage\hfill
\minipage{0.495\textwidth}
\includegraphics[width=\linewidth,clip=]{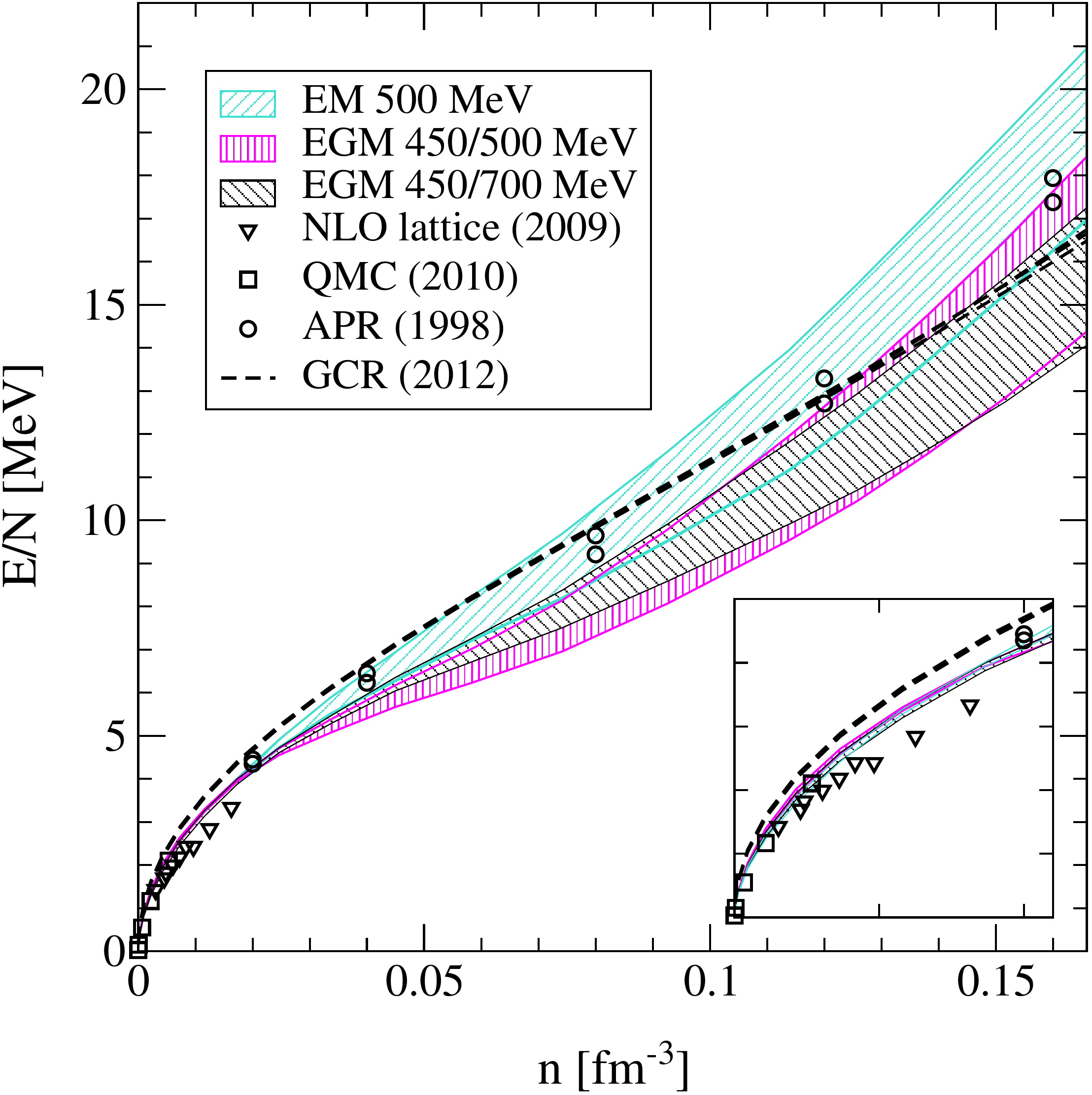}
\endminipage}
\end{center}
\caption{Energy per particle $E/N$ of neutron matter as a function 
of density $n$ based on different chiral EFT interactions and using
different many-body methods The uncertainty bands in the left panel
show the energy range based on the 500~MeV N$^3$LO NN potential of
Reference~\cite{Ente03EMN3LO} and including N$^2$LO 3N forces in MBPT
(red lines)~\cite{Hebe13ApJ} or in the SCGF
approach~\cite{Carb14SCGF}, as well as including all 3N and 4N
interactions to N$^3$LO~\cite{Tews13N3LO,Krue13N3LOlong} (cyan
band). The blue band shows the results after RG-evolution of the NN
potential~\cite{Hebe10nmatt,Hebe13ApJ}. In addition, we show results
obtained in CC theory~\cite{Hage13ccnm} and in MBPT of Corragio et
al.~\cite{Cora12neutmat}. When bands are given, these are dominated by
the uncertainties in the $c_i$ couplings in 3N forces. Figure adapted
from Reference~\cite{Hebe13ApJ}.  The right panel shows the energy per
particle including NN, 3N and 4N forces at N$^3$LO based on different
N$^3$LO potentials (cyan, magenta, and black bands). The bands include
uncertainty estimates due to the many-body calculation, the $c_i$
couplings, and by varying the 3N/4N cutoffs. For details
see~\cite{Tews13N3LO,Krue13N3LOlong}. For comparison, results are
shown at low densities (see also the inset) from NLO
lattice~\cite{Epel08lattice} and quantum Monte Carlo (QMC)
simulations~\cite{Geze09neutmat}, and at nuclear densities from
variational (APR)~\cite{Akma98EOS} and auxiliary field diffusion Monte
Carlo calculations (GCR)~\cite{Gand11nm} based on 3N potentials
adjusted to nuclear matter properties.\label{fig:nm}}
\end{figure}

The left panel of \textbf{Figure~\ref{fig:nm}} shows the energy per
particle of neutron matter up to saturation density $n_0 = 0.16 \,
{\rm fm}^{-3}$. The results are obtained with different many-body
methods based on chiral EFT interactions, all with the 500~MeV N$^3$LO
potential~\cite{Ente03EMN3LO} at the NN level. In case where the
energy is shown with bands, for MBPT (red lines, cyan and blue
band)~\cite{Hebe13ApJ,Tews13N3LO,Krue13N3LOlong} and
SCGF~\cite{Carb14SCGF} results, the theoretical uncertainty of the
energy is dominated by uncertainties in the low-energy couplings $c_1$
and $c_3$, which determine the long-range two-pion-exchange parts of
3N forces, and not by truncations in the many-body calculation. The
red lines and blue band show results including contributions from
N$^2$LO 3N forces, whereas the cyan band includes all 3N and 4N
interactions to N$^3$LO. In addition, for the blue band the NN
potential has been RG-evolved to a low-momentum scale $\Lambda = 2.0
\, {\rm fm}^{-1}$. We also show CC~\cite{Hage13ccnm}, MBPT of
Reference~\cite{Cora12neutmat}, and SCGF~\cite{Carb14SCGF} results,
including N$^2$LO 3N forces. These all lie within the overlap of the
blue and cyan band (except for the lowest density SCGF point,
where the zero-temperature extrapolation may be difficult).
The determination of the $c_i$ couplings from $\pi N$ scattering is
consistent with the extraction from NN scattering, see, e.g., the
discussion in~\cite{Hamm12RMP}, however with large uncertainties.
Therefore, the $c_i$ range for the bands in Figure~\ref{fig:nm} is
taken conservatively: at N$^2$LO (red lines and blue band), $c_1 =
−(0.7-1.4) \, {\rm GeV}^{-1}$ and $c_3 = −(2.2-4.8) \, 
{\rm GeV}^{-1}$~\cite{Hebe10PRL,Hebe13ApJ} [with a similar $c_i$ range
for SCGF~\cite{Carb14SCGF}], and at N$^3$LO (cyan band), $c_1 =
-(0.75-1.13) \, {\rm GeV}^{-1}$ and $c_3 =-(4.77-5.51) \, 
{\rm GeV}^{-1}$~\cite{Kreb123Nlong}.

Figure~\ref{fig:nm} shows that chiral EFT interactions provide strong
constraints on the EOS of neutron matter, which are consistent among
different many-body methods and considering variations of the
Hamiltonian. The remarkable overlap of the red lines and the blue band
indicates that neutron matter is, to a good approximation,
perturbative for chiral NN interactions with $\Lambda \lesssim 500 \,
{\rm MeV}$, see Reference~\cite{Krue13N3LOlong} for details.  This has
been benchmarked by first quantum Monte Carlo calculations with local
chiral EFT interactions~\cite{Geze13QMCchi,Geze14long}. In addition,
there are calculations of neutron matter using in-medium chiral
perturbation theory approaches with similar
results~\cite{Holt13PPNP,Laco11matt}.

\begin{figure}[t]
\begin{center}
\includegraphics[width=0.525\textwidth,clip=]{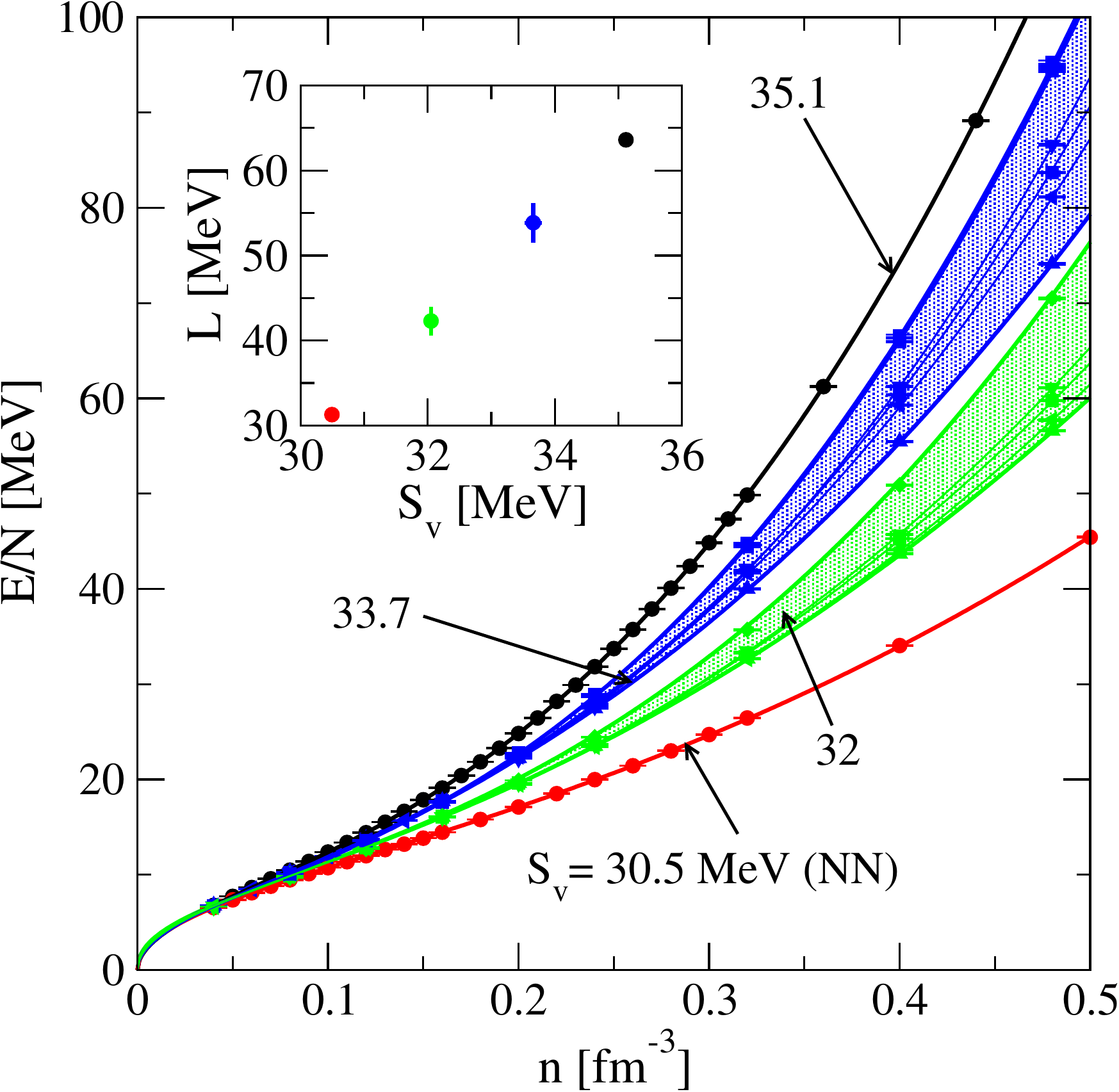}
\end{center}
\caption{Energy per particle $E/N$ of neutron matter as a function 
of density $n$ obtained from quantum Monte Carlo calculations with
phenomenological NN and 3N potentials~\cite{Gand11nm}, where the
strength of the short-range 3N interactions is adjusted to the chosen
values for the symmetry energy $S_v$. ($S_v = 30.5 \, {\rm MeV}$ is
the symmetry energy obtained for this NN potential.) For each value of
$S_v$, the band shows contributions of different 3N operator structures.
The inset illustrates a correlation between $S_v$ and the $L$
parameter. For details see~\cite{Gand11nm}.\label{fig:Gandnm}}
\end{figure}

The right panel of Figure~\ref{fig:nm} shows the first complete
N$^3$LO calculation of the neutron matter energy, which includes all
NN, 3N and 4N interactions to N$^3$LO~\cite{Tews13N3LO,Krue13N3LOlong}.
The energy range is based on different NN potentials, a variation of
the $c_i$ couplings (which dominates the total uncertainty), a
3N/4N-cutoff variation, and the uncertainty in the many-body
calculation. We note that the individual 3N topologies at N$^3$LO (see
Figure~\ref{fig:diag_N3LO}) give significant contributions to the
energy~\cite{Tews13N3LO,Krue13N3LOlong}. The N$^3$LO range in the
right panel of Figure~\ref{fig:nm} is in very good agreement with NLO
lattice results~\cite{Epel08lattice} and quantum Monte Carlo
calculations~\cite{Geze09neutmat} at very low densities (see inset),
where the properties are determined by the large scattering length and
effective range~\cite{Schw05dEFT}.  We also find a very good agreement
with other ab initio calculations of neutron matter based on the
Argonne NN and Urbana 3N potentials: The results based on variational
calculations (APR) \cite{Akma98EOS} are within the N$^3$LO band. In
addition, the results from auxiliary field diffusion Monte Carlo
calculations (GCR)~\cite{Gand11nm} are shown based on nuclear force
models adjusted to an energy difference of $32 \mev$ between neutron
matter and the empirical saturation point, see
\textbf{Figure~\ref{fig:Gandnm}}.

The properties of neutron matter impact the neutron distributions in
nuclei. In particular, a higher neutron matter pressure at typical
nuclear densities pushes neutrons further out and thus implies larger
neutron skins~\cite{Brown00radii,Type01radii}. Using these
correlations the neutron matter results shown in Figure~\ref{fig:nm}
(blue bands in the left panel) predict the neutron skin of $^{208}$Pb
to $0.17 \pm 0.03 \, {\rm fm}$~\cite{Hebe10PRL}. This is in excellent
agreement with the extraction of $0.156\substack{+0.025 \\ -0.021} \,
{\rm fm}$ from the dipole polarizability~\cite{Tami11dipole}. The
theoretical uncertainty is also smaller than the target goal of a new
PREX measurement using parity violating electron scattering at
JLAB~\cite{Abra12prex}. Moreover, including properties of doubly magic
nuclei as constraints, in addition to low-density neutron matter
results, leads to even tighter predictions for the neutron skins of
$^{208}$Pb and $^{48}$Ca to be $0.182 \pm 0.010 \, {\rm fm}$ and
$0.173 \pm 0.005 \, {\rm fm}$, respectively~\cite{Brow14skyrme}.

\subsection{Symmetry energy}
\label{symen}

The symmetry energy characterizes the behavior of the energy of
nuclear matter as a function of proton fraction $x = n_p/n$ or
asymmetry parameter $\beta = (N-Z)/A = (1 - 2 x)$. Around symmetric
matter, the energy per particle can be expanded in the following form:
\begin{equation}
\frac{E(x,n)}{A} = \frac{E(0,n)}{A} + \beta^2 E_{\text{sym}}(n) + \ldots \,.
\end{equation}
Since the symmetry energy is a key quantity for the equation of state
and for astrophysical applications, it has been subject of many
experimental and theoretical studies~\cite{Tsan12esymm,Latt12esymm,Li14esymm}.
Sometimes the above expansion is truncated after the quadratic term,
however, in general the symmetry energy $S_v$ is defined as the second
derivate with respect to proton fraction,
\begin{equation}
S_v = \frac{1}{8} \frac{\partial^2 E/A(\bar{n},x)}{\partial x^2} 
\biggr|_{\bar{n}=1, x=1/2},
\end{equation}
where $\bar{n} = n/n_0$ is the density in units of the saturation
density. Another important quantity for astrophysics is the density
dependence of the symmetry energy, characterized by the $L$ parameter,
\begin{equation}
L = \frac{3}{8} \frac{\partial^3 E/A(\bar{n},x)}
{\partial \bar{n} \partial x^2} \biggr|_{\bar{n}=1, x=1/2}.
\end{equation}

\begin{table}[t]
\caption{Predicted range for the symmetry energy $S_v$ and the $L$ 
parameter, which determines the density dependence of the symmetry
energy. The results~\cite{Hebe13ApJ} are
obtained from neutron matter calculations based on chiral NN+3N
forces using the expansion Equation~\ref{eq:Eskyrme} with
different $\gamma$ values, which lead to different incompressibilities
$K$. As shown, the predicted ranges for $S_v$ and $L$ depend very
weakly on $\gamma$. Also given is the predicted range for the $L$
parameter based on quantum Monte Carlo calculations with
phenomenological NN+3N potentials, where the strength of the
short-range 3N interactions was adjusted to the chosen range for
$S_v$~\cite{Gand11nm} (see also Figure~\ref{fig:Gandnm}).\label{tab:SvL}}
\begin{center}
\begin{tabular}{@{}lc|c|c@{}}
$\gamma$ & $K$ [MeV] & $S_v$ [MeV] & $L$ [MeV] \\
\hline
$1.2$ & 210 & $29.7 - 32.8$ & $32.4 - 53.4$ \\
$4/3$ & 236 & $29.7 - 33.2$ & $32.5 - 57.0$ \\
$1.45$ & 260 & $30.1 - 33.5$ & $33.6 - 56.7$ \\
\hline
\multicolumn{2}{@{}l|}{Gandolfi et al. (2011)} & $32.0 - 35.1$ & $40.6 - 63.6$
\end{tabular}
\end{center}
\end{table}

To obtain $S_v$ and $L$ it is necessary to extend the microscopic
calculations to finite proton fractions. This can be achieved by
either performing calculations for several different asymmetries or by
using neutron matter and symmetric nuclear matter as anchor points and
interpolating these results to arbitrary proton fractions. One such
interpolation has been used in Reference~\cite{Hebe13ApJ}, with an
empirical parameterization that includes kinetic energy terms plus an
interaction energy that is quadratic in the neutron excess $1 - 2x$:
\begin{eqnarray}
\frac{E/A(\bar{n},x)}{T_0} &=&
\frac{3}{5} \left[ x^{5/3} + (1-x)^{5/3} \right] (2 \bar{n})^{2/3} 
\nonumber \\[2mm]
&&- \left[ ( 2 \alpha - 4 \alpha_L) x (1 - x) + \alpha_L \right] \bar{n} 
+ \left[ ( 2 \eta - 4 \eta_L) x (1 - x) + \eta_L \right]
\bar{n}^{\gamma} \,. \label{eq:Eskyrme}
\end{eqnarray}
Here, $T_0 = 36.84 \, \rm{MeV}$ is the Fermi energy of symmetric
nuclear matter at the saturation density. The parameters $\alpha,
\eta, \alpha_L$, and $\eta_L$ can be determined from the empirical
saturation properties of symmetric nuclear matter combined with
microscopic calculations of neutron matter~\cite{Hebe13ApJ}. This
strategy takes advantage of the fact that the theoretical
uncertainties of neutron matter calculations are significantly smaller
than for systems with finite proton fractions. As shown in
\textbf{Table~\ref{tab:SvL}}, the predicted range for $S_v$ and $L$
depends only weakly on the particular choice of $\gamma$, which is
correlated with the incompressibility. Also given is the predicted
range for the $L$ parameter obtained from quantum Monte Carlo
calculations~\cite{Gand11nm} based on 3N potentials adjusted to the
chosen range of $S_v$. Figure~\ref{fig:Gandnm} shows these results for
neutron matter in more detail, highlighting the correlation between
$S_v$ and $L$. It is remarkable how well the $S_v - L$ regions in
Table~\ref{tab:SvL} agree, given the very different Hamiltonians used
in the quantum Monte Carlo calculations.

The $S_v - L$ predictions based on NN+3N forces agree well with
constraints obtained from energy density functionals for nuclear
masses~\cite{Kort10edf} and from the $^{208}$Pb dipole
polarizability~\cite{Tami11dipole}. In addition, there is good
agreement with studies of the Sn neutron skin~\cite{Chen10skin}, of
giant dipole resonances~\cite{Trip08gdr}, and with an estimate
obtained from modeling X-ray bursts and quiescent low-mass X-ray
binaries~\cite{Stei10eos}. A detailed discussion can be found in
Reference~\cite{Latt12esymm}.

\begin{figure}[t]
\begin{center}
\includegraphics[width=0.525\textwidth,clip=]{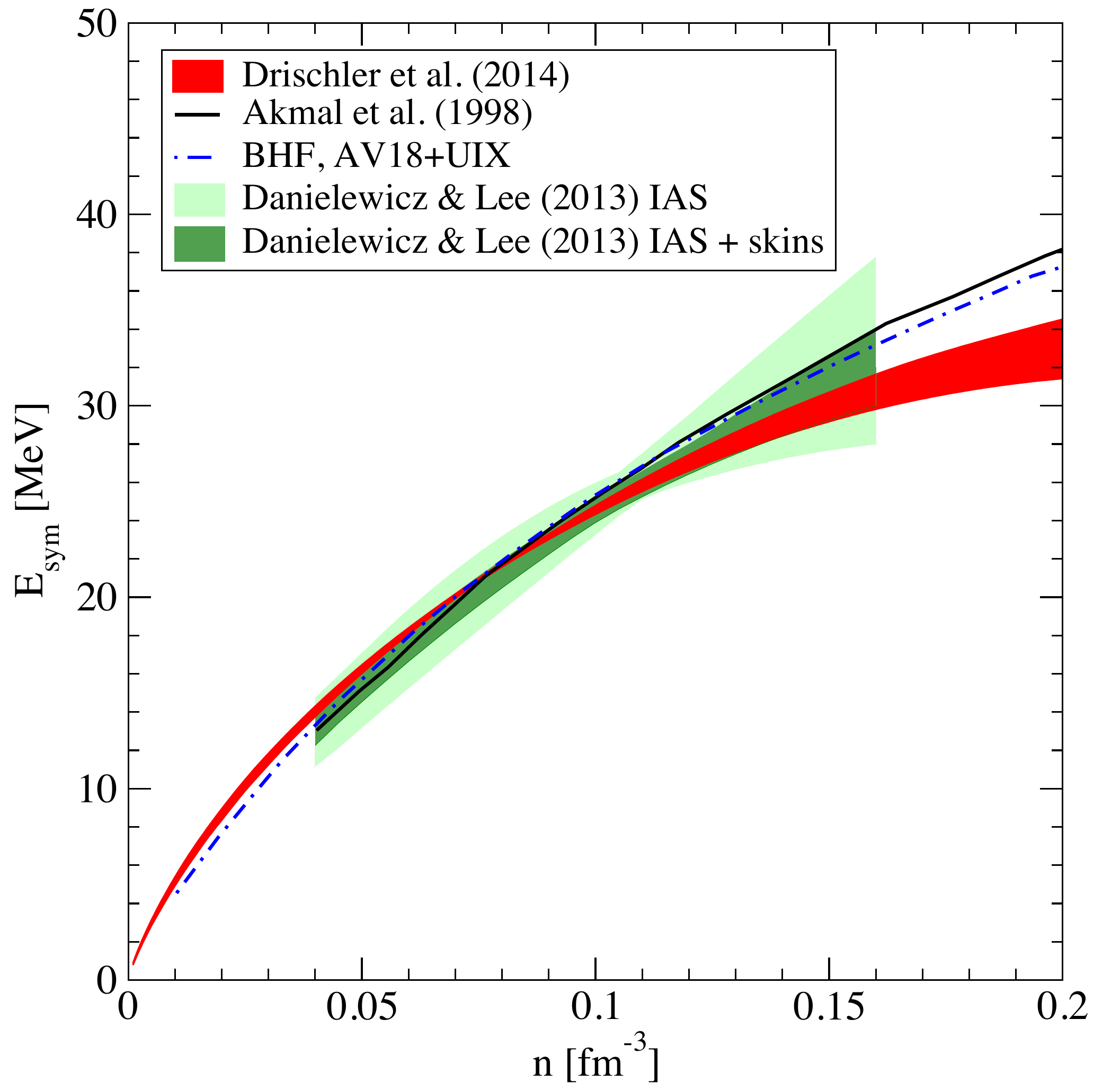}
\end{center}
\caption{Symmetry energy as a function of density $n$ obtained from 
ab initio calculations of asymmetric matter based on chiral NN+3N 
forces~\cite{Dris14asymmat} (red band), from microscopic
calculations performed with a variational approach [Akmal et al.
(1998)]~\cite{Akma98EOS} and from Brueckner-Hartree-Fock calculations
(BHF)~\cite{Tara13bhf} based on the Argonne $v_{18}$ NN and Urbana UIX
3N potentials (with parameters adjusted to the empirical saturation
point). For comparison, the band over the density range $n = 0.04-0.16 \,
{\rm fm}^{-3}$ is based on an analysis of isobaric analog states
(IAS) and including the constraints from neutron skins (IAS +
skins)~\cite{Dani13esymm}. Figure taken from~\cite{Dris14asymmat}.
\label{fig:Esymm}}
\end{figure}

Recently, the symmetry energy has also been studied in ab initio
calculations of asymmetric matter at small proton fractions based on
chiral EFT interactions~\cite{Dris14asymmat}. The energy of asymmetric
matter was found to compare very well with the quadratic expansion
even for neutron-rich conditions, which was then used to extract the
quadratic symmetry energy term $E_{\rm sym}$. In addition, the results
were used to benchmark the empirical parameterization,
Equation~\ref{eq:Eskyrme}, with very good agreement. In contrast to
other calculations, the results are based on 3N forces fit only to
light nuclei, without adjustments to empirical nuclear matter
properties. The results for $E_{\rm sym}$ are compared in
\textbf{Figure~\ref{fig:Esymm}} with constraints from a recent
analysis of isobaric analog states (IAS) and including the constraints
from neutron skins (IAS + skins)~\cite{Dani13esymm}, showing a
remarkable agreement over the entire density range. Compared to
extracting the symmetry energy from neutron matter calculations using
the empirical parameterization, Equation~\ref{eq:Eskyrme}, the
uncertainty is reduced due to the explicit information from asymmetric matter.

\subsection{Neutron stars}
\label{nstar}

The structure of nonrotating neutron stars can be studied by solving
the Tolman-Oppenheimer-Volkov equations~\cite{Oppe39tov}, based on an
EOS of neutron star matter, i.e., matter in beta-equilibrium. The
proton fraction $x$ for matter under these conditions is determined by
minimizing, for a given nucleon density, the total energy per particle
plus the contributions from electrons and from the rest mass of the
nucleons. The resulting proton fractions are of the order $x \approx 5
\%$~\cite{Hebe10PRL,Hebe13ApJ}.

Since the central densities of neutron stars can significantly exceed
the regime for which reliable microscopic nuclear matter calculations
are possible, it is necessary to extend the EOS systematically to
higher densities. This can be achieved by using microscopic results up
to a maximal density and employing general piecewise polytropic
extensions beyond~\cite{Hebe10PRL,Hebe13ApJ}. This strategy allows
for soft regions and generates a complete set of possible EOSs at high
densities, independent of the assumptions on the interactions and
constituents of matter at high densities. In the end only those EOS
are retained that (a) remain causal for all relevant densities, and
(b) are able to support a neutron star mass $M = \widehat{M}$, the
mass of the heaviest observed neutron star.

\begin{figure}[t]
\begin{center}
{\minipage{0.495\textwidth}
\includegraphics[width=\linewidth,clip=]{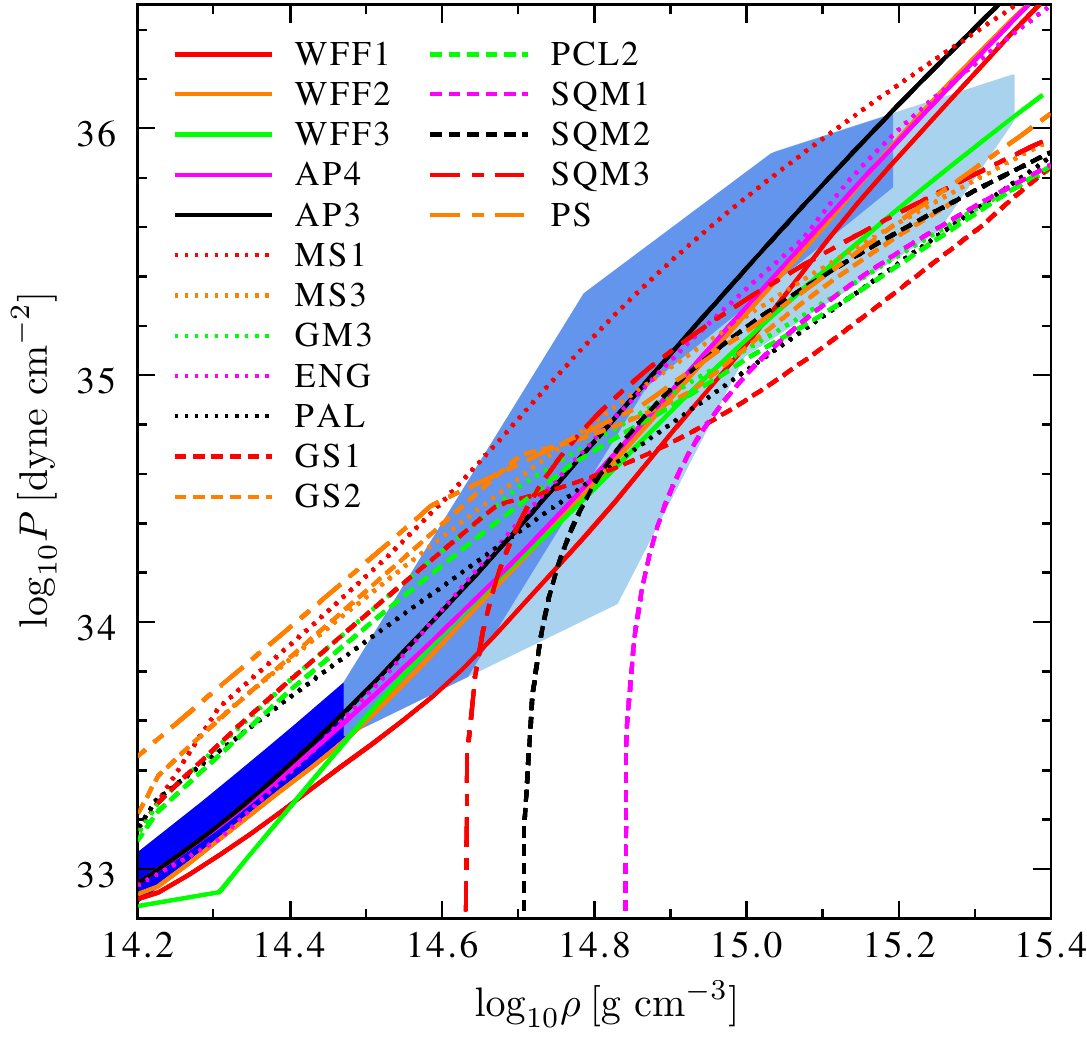}
\endminipage\hfill
\minipage{0.495\textwidth}
\includegraphics[width=\linewidth,clip=]{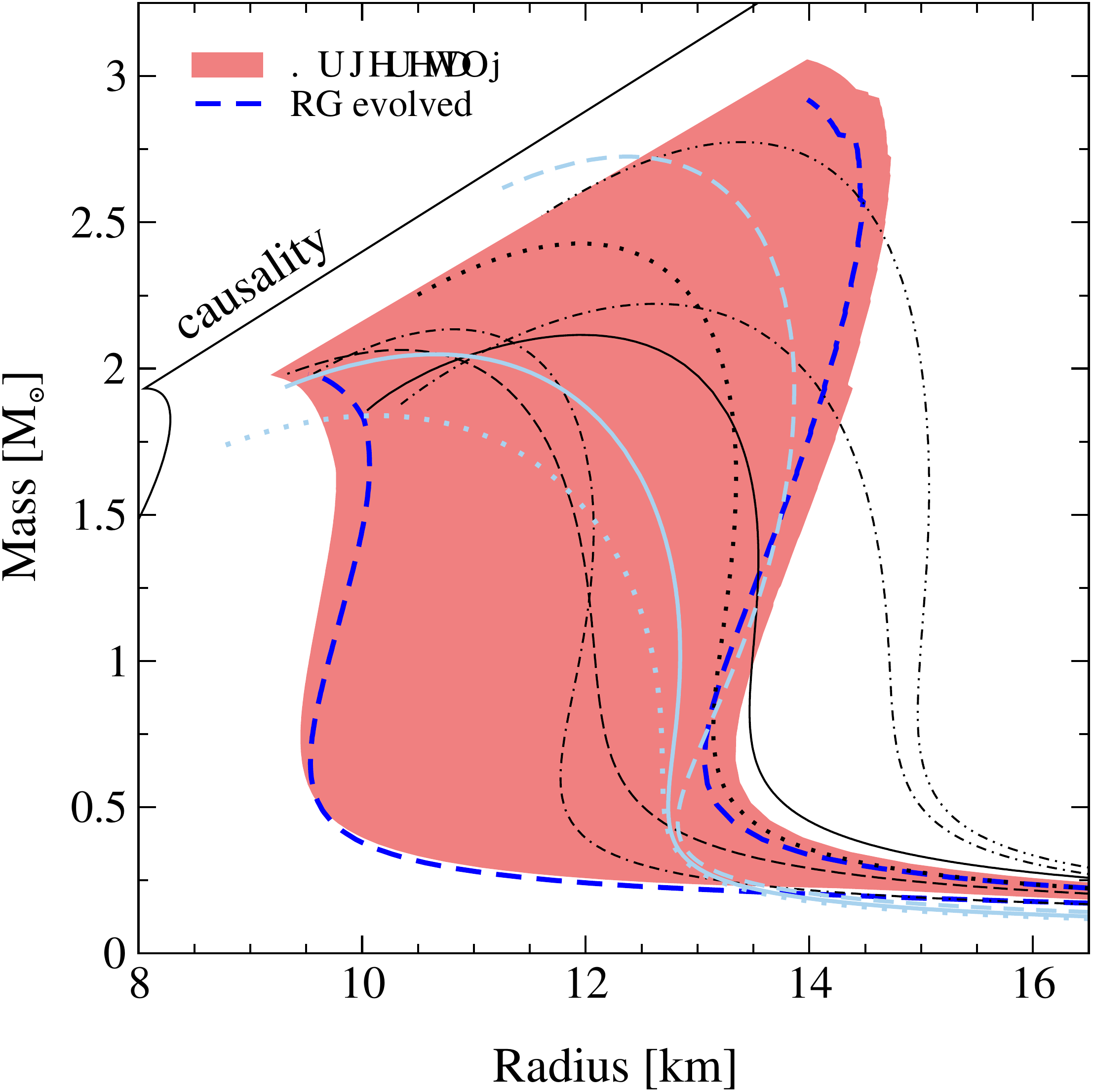}
\endminipage}
\end{center}
\caption{Left panel:~Constraints for the pressure $P$ of neutron star
matter as a function of mass density $\rho$ compared to EOSs commonly
used to model neutron stars~\cite{Latt00ns}. The blue band at low
densities represents the pressure predicted by the neutron matter
results of Figure~\ref{fig:nm} (red lines of the left panel)
and incorporating beta equilibrium.  The bands at high densities are
the envelope of general polytropic extensions that are causal and
support a neutron star of mass $\widehat{M} = 1.97 \, M_\odot$ (light
blue band) and $\widehat{M} = 2.4 \, M_\odot$ (darker blue band). For
details see~\cite{Hebe13ApJ}. Right~panel:~Constraints on
the mass-radius diagram of neutron stars based on the N$^3$LO neutron
matter results of Figure~\ref{fig:nm} (right panel) and
following~\cite{Hebe13ApJ} for the extension to neutron-star
matter and to high densities (red band), in comparison to the
constraints based on RG-evolved NN interactions and 3N forces at
N$^2$LO (thick dashed blue lines, based on the $1.97 \,
M_\odot$ band of the left panel). We also show the mass-radius
relations obtained from EOSs for supernova simulations.
For details see~\cite{Krue13N3LOlong}.\label{fig:nstar}}
\end{figure}

The left panel of \textbf{Figure~\ref{fig:nstar}} shows the resulting
uncertainty band for the pressure as a function of mass density. The
blue band at lower densities represents the pressure predicted for
matter in beta equilibrium based on chiral EFT interactions (see red
dashed lines in the left panel of Figure~\ref{fig:nm}). The bands at
higher densities give the EOS range, which is the envelope of all
allowed polytropes at higher densities. The lighter blue band at high
densities corresponds to the mass constraint $\widehat{M} = 1.97
M_{\odot}$, the central value of the two-solar-mass neutron star
measured by Shapiro delay~\cite{Demo10ns} and the lower $1\sigma$ mass
of the recently observed most massive neutron star from radio timing
observations~\cite{Anto13PSRM201}, whereas the darker blue band
corresponds to $\widehat{M} = 2.4 M_{\odot}$, a fictitious heavier
neutron star.  Obviously, the higher the mass of the heaviest neutron
star observed, the stronger the EOS band is constrained. This
uncertainty band is compared with a representative set of EOSs used in
the literature. This set contains EOSs calculated within diverse
theoretical approaches and based on different degrees of freedom. For
details and notation we refer to Reference~\cite{Latt00ns}. We find
that a significant number of EOSs are not compatible with the lower
density band based on chiral EFT interactions. In addition, at higher
densities only very few EOSs, including the variational EOSs based on
phenomenological nuclear potentials~\cite{Akma98EOS} AP3 and AP4 in
the left panel of Figure~\ref{fig:nstar}, are within the uncertainty
bands over the entire density range. Finally, these constraints imply
that a $1.4 \, (1.97) \, M_\odot$ neutron star does not exceed
densities beyond $4.4 \, (7.6) \, n_0$, which corresponds to a Fermi
momentum of only $550 \, (660) \, {\rm MeV}$.

The EOS bands directly translate into constraints for the radii of
neutron stars. In the right panel of Figure~\ref{fig:nstar} we present
the radius constraints obtained from the EOS bands shown in the left
panel for the case $\widehat{M} = 1.97 \, M_{\odot}$ (blue dashed
lines) and the corresponding bands based on complete N$^3$LO
calculations shown in the right panel of Figure~\ref{fig:nm} (red
band). The radius uncertainty band represents an envelope of a large
number of individual EOS reflecting the uncertainties at low densities
and in the polytropic extensions to high
densities~\cite{Hebe13ApJ,Krue13N3LOlong}.The results of
Figure~\ref{fig:nstar} predict a radius range of $9.7-13.9 \,
{\rm km}$ for a $1.4 \, M_\odot$ neutron star based on the complete
N$^3$LO calculations. The largest supported neutron-star mass is found
to be $3.1 \, M_\odot$, with a corresponding radius of about $14 \,
\text{km}$.  These results agree very well with the mass-radius
constraints from the neutron matter calculations based on RG-evolved
NN interactions with N$^2$LO 3N forces. Furthermore, the radius
constraints are also consistent with astrophysical extractions
obtained from modeling X-ray burst sources, see, e.g.,
Reference~\cite{Stei10eos}.

For astrophysical applications it is crucial to reduce the theoretical
uncertainties of the EOS over the entire density range shown in the
left panel of Figure~\ref{fig:nstar}. At lower densities this requires
improved estimates and a reduction of the theoretical uncertainties in
nuclear forces. Specifically, this involves the inclusion of
higher-order contributions in the chiral expansion, systematic
order-by-order convergence studies, as well as improved determinations
of the low-energy couplings. At higher densities, novel observations
are expected to provide new and model-independent constraints: On the
one hand, the observation of heavier neutron stars leads to a
systematic reduction of the uncertainties, as illustrated in the left
panel of Figure~\ref{fig:nstar}. On the other hand, astrophysical
information on neutron star radii will provide significant constraints
on the EOS. In particular, the gravitational wave signal from mergers
of binary neutron
stars~\cite{Ande09nsgw,Baus11gw,Baus12gweos,Taka14mergers} and neutron
star-black hole mergers~\cite{Lack11bheos} has been shown to be
sensitive to properties of the EOS at high densities and to neutron
star radii. Hence, a detection with, e.g., advanced
LIGO~\cite{ligo14advligo} will significantly improve our knowledge of
the EOS in this regime. Moreover, future observations with the Neutron
star Interior Composition ExploreR (NICER) and the Large Observatory
for X-Ray Timing (LOFT) offer completely new perspectives for
measuring neutron star radii. This will in turn provide important
insights into nuclear forces at neutron-rich extremes.

\section{SUMMARY AND OUTLOOK}
\label{summary}

We have shown that the physics of 3N forces connects neutron-rich
nuclei with neutron-rich matter in neutron stars. The main features
discussed in this review can be summarized by 3N forces having two
effects for neutron-rich systems: first, 3N forces provide a repulsive
central interaction, which drives saturation and leads to a stiffening
of the neutron matter equation of state with increasing density. The
same repulsion is important for the contributions from 3N forces to
the ground-state energies that are key for the location of the neutron
dripline. Second, 3N forces provide an attractive spin-orbit
interaction, which increases the spin-orbit splittings and gives rise
to the associated magic numbers, e.g., at $N=28$. This physics
and its interplay with the repulsive central interaction is at play
for the formation and evolution of shell structure and for the
spectroscopy of neutron-rich nuclei discussed in this review. The
experimental discoveries for neutron-rich oxygen and calcium isotopes
show that 3N forces provide an exciting link between the theoretical
frontier in effective field theories and many-body methods with the
exploration of exotic nuclei at rare isotope beam facilities
worldwide. Three-nucleon forces are also key for the properties of
neutron-rich matter at nuclear densities, which impacts the neutron
skin, the symmetry energy, and the structure of neutron stars, in
particular their radii.

Finally, we list a number of open problems and opportunities. First, it is
important to study the order-by-order convergence in chiral EFT, and to carry
out first complete N$^3$LO calculations of nuclei. This is especially
important, because the $Q^4$ contributions are known to be important for an
accurate reproduction of NN phase shifts at energies relevant to nuclei. It is
crucial to study the theoretical uncertainties due to the truncation in chiral
EFT, due to uncertainties in the low-energy couplings in NN and 3N forces, and
due to the many-body calculation. While the latter has been well documented
for many approaches and with benchmarks (see, e.g., the remarkable agreement
for the same Hamiltonian with different ab initio methods discussed here), the
theoretical uncertainties in the Hamiltonian have been less explored. Because
most calculations so far are based on the N$^3$LO potential of
Reference~\cite{Ente03EMN3LO}, it is also necessary to explore different NN
potentials and many new developments in chiral EFT, including the optimized
potentials of References~\cite{Ekst14GT2bc,Carl15sim}, new local chiral
potentials~\cite{Geze14long}, and improved chiral potentials of
Reference~\cite{Epel15improved}, all with corresponding 3N forces. In
addition, it is important to explore improved power counting schemes (e.g.,
see the discussion in Reference~\cite{Nogg05renorm}) and to investigate
nuclear forces with explicit $\Delta$ degrees of
freedom~\cite{Kreb07Deltas,Piar14Deltas}.  Moreover, the discussed results
show that additional work is needed to quantify the theoretical uncertainties,
especially due to the truncation in chiral EFT \cite{Furn15uncert}, which is
particularly relevant to the long-range parts of 3N forces due to the large
$c_i$ values.

Important advances for ab initio calculations are the investigation of
open-shell nuclei with a range of many-body methods, the inclusion of
the continuum for loosely bound and resonant states close to the
dripline~\cite{Mich09GSM,Hage12Ox3N}, the study of electroweak
transitions with effective operators and consistent two-body currents
based on chiral
EFT~\cite{Koll09curr,Mene110nbb2bc,Past13momM12b,Ekst14GT2bc},
extending the calculations to heavier nuclei with a controlled
convergence in terms of 3N matrix elements included, and the reduction
of uncertainties in the equation of state to further constrain the
properties of neutron-rich matter and neutron-stars. Finally, it would
be very interesting to transport the knowledge and constraints from
nuclear forces to density functional calculations of all nuclei,
especially regarding the structures of 3N forces and their impact on
neutron-rich systems.



\section*{DISCLOSURE STATEMENT}

The authors are not aware of any affiliations, memberships, funding,
or financial holdings that might be perceived as affecting the
objectivity of this review.

\section*{ACKNOWLEDGMENTS}

We thank T.~Aumann, C.~Barbieri, K.~Blaum, S.~K.~Bogner, B.~A.~Brown, 
A.~Carbone, J.~Dilling, C.~Drischler, T.~Duguet, E.~Epelbaum, B.~Friman,
R.~J.~Furnstahl, S.~Gandolfi, A.~Gezerlis, G.~Hagen, H.-W.~Hammer, 
H.~Hergert, J.~W.~Holt, H.~Krebs, T.~Kr\"uger, J.~M.~Lattimer, A.~Nogga, 
T.~Otsuka, T.~Papenbrock, C.~J.~Pethick, A.~Poves, R.~Roth, J.~Simonis,
V.~Som\`a, O.~Sorlin, T.~Suzuki, and I.~Tews for useful discussions
on the topics of this review. This work was supported in part by the
ERC Grant No. 307986 STRONGINT, the BMBF under Contract No. 06DA70471,
the DFG through Grant SFB 634, the Helmholtz Alliance HA216/EMMI, the
National Research Council of Canada, and by an International Research
Fellowship of the Japan Society for the Promotion of
Science. Computing time at the John von Neumann Institute for
Computing in J\"ulich is gratefully acknowledged.

\bibliography{../../strongint/strongint}
\bibliographystyle{ar-style5.bst}

\end{document}